\documentclass[twocolumn,trackchanges]{aastex631}

\newcommand\emin{\epsilon_{\rm min}}

\newcommand\gammae{\gamma_{\rm e}}

\newcommand\me{m_{\rm e}}
\newcommand\sT{\sigma_{\rm T}}
\newcommand\rg{r_{\rm g}}

\newcommand\Er{E_{\rm r}}
\newcommand\Bh{B_{\rm H}}

\newcommand\Lcur{L_{\rm cur}}

\newcommand\Lbz{L_{\rm BZ}}
\newcommand\rh{r_{\rm H}}
\newcommand\kb{k_{\rm B}}

\newcommand\nism{n_{\rm ISM}}
\newcommand\m{m_{p}}

\newcommand\rin{r_{\rm in}}
\newcommand\ro{r_{\rm out}}
\newcommand\rnull{r_{\rm null}}
\newcommand\rhoGJ{\rho_{\rm GJ}}
\newcommand\epsilons{\epsilon_{\rm s}}
\newcommand\tcic{t_{\rm cool}^{\rm ic}}
\newcommand\tccur{t_{\rm cool}^{\rm cur}}
\newcommand\tdyn{t_{\rm dyn}}

\revised{Jan 15, 2024}
\accepted{Jan 18, 2024}

\shorttitle{Gamma-rays from stellar-mass BH magnetospheric gaps}
\shortauthors{Kin et al.}
\graphicspath{{./}{figures/}}
\usepackage{amsmath, amssymb}
\usepackage{comment}
\usepackage{url}
\begin{document}
\title{
1D GRPIC Simulations of Stellar-Mass Black Hole Magnetospheres: Semi-Analytic Model of Gamma-Rays from Gaps}

\author[0000-0002-9712-3589]{Koki Kin}
\affiliation{Astronomical Institute, Graduate School of Science, Tohoku University, Sendai 980-8578, Japan}

\author[0000-0002-2498-1937]{Shota Kisaka}
\affiliation{Physics Program, Graduate School of Advanced Science and Engineering, Hiroshima University, Higashi-Hiroshima, 739-8526, Japan}

\author[0000-0002-7114-6010]{Kenji Toma}
\affiliation{Astronomical Institute, Graduate School of Science, Tohoku University, Sendai 980-8578, Japan}
\affiliation{Frontier Research Institute for Interdisciplinary Sciences, Tohoku University, Sendai 980-8578, Japan}

\author[0000-0003-2579-7266]{Shigeo S. Kimura}
\affiliation{Astronomical Institute, Graduate School of Science, Tohoku University, Sendai 980-8578, Japan}
\affiliation{Frontier Research Institute for Interdisciplinary Sciences, Tohoku University, Sendai 980-8578, Japan}

\author[0000-0001-7572-4060]{Amir Levinson}
\affiliation{The Raymond and Beverly Sackler, School of Physics and Astronomy, Tel Aviv University, Tel Aviv 69978, Israel}



\begin{abstract}
In the absence of a sufficient amount of plasma injection into the black hole (BH) magnetosphere, the force-free state of the magnetosphere cannot be maintained, leading to the emergence of strong, time-dependent, longitudinal electric field 
(spark gap).
Recent studies of supermassive BH magnetospheres by using analytical methods and particle-in-cell (PIC) simulations
propose the possibility of the efficient particle acceleration and consequent gamma-ray emission in the spark gap.
In this work, we perform one-dimensional general relativistic PIC simulations to examine the gamma-ray emission from stellar-mass BH magnetospheres. We find that intermittent spark gaps emerge and particles are efficiently accelerated, in a similar manner to the supermassive BH case. We build a semi-analytic model of the plasma dynamics and radiative processes which reproduces the maximum electron energies and peak gamma-ray luminosities in the simulation results.
Based on this model, we show that gamma-ray signals from stellar-mass BHs wandering through the interstellar medium could be detected by gamma-ray telescopes such as the Fermi Large Area Telescope, or the Cherenkov Telescope Array. 
\end{abstract}



\section{Introduction} \label{sec:intro}
Relativistic jets from gas-accreting black holes (BHs) are widely believed to be activated via the electromagnetic extraction of the BH's rotational energy, i.e. Blandford-Znajek (BZ) process \citep{BZ77}.
General relativistic magnetohydrodynamic simulations show that the highly-magnetized region, i.e. the magnetosphere, can be formed around a BH, where the BZ process works (e.g. \citealp{Mckinney04, Tchekho11, Narayan12, Takahashi16, Nakamura18,Porth19,Mizuno21}).
The viability of the BZ process under general conditions is also supported by analytical studies (\citealp{Komissa04}; \citealp{Toma14}; \citealp{Kimura_M21}; \citealp{Komissa22}, and references therein).

Continuous injection of plasma into the magnetosphere is required to activate the steady force-free jet.
A possible injection process is electron-positron ($e^+e^-$) pair productions by ambient MeV energy photons from the accretion disk surrounding the magnetosphere, but it could be insufficient to maintain the force-free state in the case of a low accretion rate (\citealp{Beskin92, Hirotani98, Levinson11, Moscibro11, Hirotani16, Wong21, Yao21}; see \citealp{Kimura20} and \citealp{Kuze22} for a possible contribution of GeV photons).
Magnetic reconnections around the equatorial plane close to the BH may also cause $e^+e^-$ pair loading \citep{Ripperda22, Kimura22, Chen22, Hakobyan23} or direct injection of accreting plasma. 
However, the former mechanism can inject large amount of plasma only during a flaring state, and the latter mechanism may not screen the entire region of the magnetosphere \citep{Niv23}.

In the absence of sufficient plasma injection, the magnetosphere becomes charge-starved, giving rise to longitudinal electric fields i.e. spark gaps.
The charged particles are linearly accelerated in the gap, and these scatter the ambient soft photons to the gamma-ray energy, which create additional $e^+e^-$ pairs.
The nature of gaps in the BH magnetosphere has been investigated mainly for supermassive BHs (SMBHs) 
in several literatures, including analytic models \citep{Beskin92, Hirotani98, Hirotani16, Hirotani17, Levinson17, Katsoulakos20} and numerical studies with general relativistic particle-in-cell (GRPIC) simulations of 1D local gaps \citep{Levinson18, Chen20, Kisaka20, Kisaka22} and 2D global magnetospheres \citep{Parfrey19, Crin20, Crin21, Niv23, Hirotani22, Hirotani23}.\footnote{
\citet{Chen20}, \citet{Hirotani22}, and \citet{Hirotani23} performed GRPIC simulations for the stellar-mass BH system, but they do not discuss on the gamma-ray signals in detail.}
These studies indicate that the gamma-ray emission escaping from the system could be bright enough to detect and could explain the observed rapid GeV-TeV flares from active galactic nuclei (AGNs), including BL Lacs (PKS 2155-304 \citealp{Aha07}, Markarian 501 \citealp{Albert07}, BL Lac \citealp{Arlen13, Magic19, Pandey22}), flat spectrum radio quasars (PKS 1222+21 \citealp{Aleksic11}, 3C279 \citealp{Ackermann16}, CTA102 \citealp{Shukla18}), and radio galaxies (M87 \citealp{Abramowski12}, IC310 \citealp{Aleksic14}).

Similar spark gaps may arise in charge-starved magnetospheres formed by accretion of interstellar medium (ISM) onto stellar-mass BHs, which are the source candidates of unidentified gamma-ray objects \citep{Hirotani18}.
The size of the system, expected magnetic field strength, and characteristics of soft photons from the accretion disk differ significantly from SMBH magnetospheres, and thus the plasma dynamics for stellar-mass BH should be examined separately.
\citet{Hirotani16_2}, \citet{Song17}, and \citet{Hirotani18} present analytic steady gap models for stellar-mass BHs, but it is unlikely that the steady solutions are realized over broad ranges of parameters \citep{Levinson17}.
A simulation-based modeling of time-dependent gaps is required to predict their emission properties.

In the present paper we examine the dynamics of a local spark gap of stellar-mass BHs and the subsequent radiation characteristics, by using 1D GRPIC simulations.\footnote{Although 2D simulations could provide more complete analysis on the global magnetospheric structure \citep[e.g.,][]{Parfrey19, Crin20, Crin21},
they cannot resolve the skin depth for realistic conditions and require proper rescaling that needs to be better assessed.}
We broadly explore the dependence of the gap activity on the ambient photon characteristics and the magnetospheric current (Section \ref{sec:simu_res}). 
Furthermore, we construct a semi-analytic model of the particle acceleration and radiation processes in the gap which reproduces the maximum particle energies and the peak gamma-ray luminosities in the simulations (Section \ref{sec:semiana}).
We also calculate the spectra of escaping gamma-rays in some cases with our semi-analytic model (Section \ref{sec:impli}) and briefly discuss the detectability (Section \ref{sec:dandc}).

\section{The Numerical Simulation} \label{sec:simu}
We use the 1D GRPIC code described in \citet{Levinson18}, \citet{Kisaka20}, and \citet{Kisaka22} (the latter two are hereafter denoted as K20 and K22, respectively).
In this code, we assume that the gap activity does not significantly affect the global magnetospheric structure, where the magnetic field is set to have the split-monopole configuration.
Under this assumption, we solve the Maxwell's equations for the longitudinal electric field $E_r$ measured in the frame of a zero angular momentum observer (ZAMO)
in a background Kerr spacetime, given in Boyer-Lindquist coordinates (see Appendix~\ref{space}),
\begin{equation}
\label{rGauss}
\partial_{\rm r}\left(\sqrt{A}\Er\right)=4\pi\Sigma\left(j^t-\rho_{\rm GJ}\right),
\end{equation}
and
\begin{equation}
\label{rAmp}
\partial_t\left(\sqrt{A}\Er\right)=-4\pi\left(\Sigma j^r-J_0\right),
\end{equation}
where $j^t$ is the charge density and $j^r$ is the current density.
The Goldreich-Julian (GJ) charge density $\rho_{\rm GJ}$ is written as
\begin{equation}
\begin{aligned}
\label{rhoGJ}
\rho_{\rm GJ}
&=\dfrac{\sqrt{A(r_{\rm H})}\Bh\cos\theta}{2\pi\Sigma^2\Delta}\left[\left(A+\dfrac{2\rg r(r^2+a^2)}{\Sigma}a^2\sin\theta\right)\right.\\
&\hspace{70pt}\times(\omega_{\rm H}-\Omega_{\rm F})+\Delta\omega_{\rm H}a^2\sin^2\theta\biggr],
\end{aligned}
\end{equation}
where $\Bh$ denotes the magnetic field strength at the horizon, $\rg = GM/c^2$ the gravitational radius of the BH with mass $M$ ($G$ is the gravitational constant), $\rh = \rg(1+\sqrt{1-a_*^2})$ the horizon radius, $\omega_\mathrm{H} = a_*c/2r_\mathrm{H}$ the angular velocity of the BH ($a_*\equiv a/\rg$ is the dimensionless spin parameter), and $\Omega_\mathrm{F}=\omega_\mathrm{H}/2$ the angular velocity of the magnetic surface.
$J_0$ is the magnetospheric current parameterized by
\begin{equation}
\label{J0}
J_0=j_0\dfrac{\omega_\mathrm{H}B_\mathrm{H}}{4\pi}(r_\mathrm{H}^2+a^2)\cos\theta.
\end{equation}
$j_0=-1$ corresponds to the magnetospheric current in the BZ force-free solution (for the counter-clockwise rotation around the rotation axis).\footnote{\citet{Levinson18}, K20, and K22 mistook the parameterization of $J_0$.
The models of $j_0=-1$ in their studies are actually the cases that $J_0$ is $1/2\pi$ times weaker than the BZ current.} 
We note that the null charge surface, where $\rhoGJ = 0$, $r = r_{\rm null}(\theta)$, is nearly spherical.

The equation of motion
for electrons and positrons is
\begin{equation}
\begin{aligned}
\label{EoM}
\dfrac{du_i}{dt}=-\sqrt{g_{\rm rr}}\gamma_i\partial_{\rm r}(\alpha)+\alpha\left(\dfrac{q_i}{\me}\Er-\dfrac{P_{\rm cur}(\gamma_i)}{\me v_i}\right),
\end{aligned}
\end{equation}
where $q_i$ is the charge of $i$th particle, $u_i=\sqrt{g_{\rm rr}}u_i^r$ the normalized four-velocity component in the $r$ direction measured by a ZAMO, $\gamma_{i}$ the Lorentz factor, $v_i=u_i/\gamma_i$ the three-velocity component, and $P_{\mathrm{cur}}(\gamma_i)=2e^2\gamma_i^4v_i/3r_\mathrm{g}^2$ the curvature power with the the curvature radius fixed as $\rg$.
For photons we solve
\begin{equation}
\label{pmom}
\dfrac{d\tilde{p}_j^r}{dt}=-\sqrt{g^{\rm rr}}\tilde{p}_j^t\partial_{\rm r}(\alpha),
\end{equation}
where $\tilde{p}_j^\mu$ ($\mu=t,r$) is the momentum components of $j$th photon measured by a ZAMO.

We consider two main processes of photon interactions by using the Monte-Carlo approach. One is the inverse Compton (IC) scattering of soft `seed' photons from the accretion disk and the other is the $\gamma\gamma$ pair production by the scattered photons and the seed photons.
We assume the intensity of the seed photon field to be steady, isotropic, and homogeneous with a simple power-law spectrum
$I_\mathrm{s}(\epsilons)=I_0(\epsilons/\emin)^{-p}\,(\epsilon_{\mathrm{min}}\leq\epsilons\leq\epsilon_{\mathrm{max}})$, where $\epsilons$ is the seed photon energy normalized by an electron rest mass energy $m_\mathrm{e}c^2$.
The fiducial optical depth $\tau_0$ is related to the intensity normalization $I_0$ as
$\tau_0=n_\mathrm{s}(\epsilon_{\mathrm{min}})\sigma_\mathrm{T}r_\mathrm{g}=4\pi\sigma_\mathrm{T}r_\mathrm{g}I_0/hc$,
where $\sT$ is the Thomson cross section and $h$ is the Planck constant.
We compute the momenta of IC scattered photons by Eq.~(\ref{pmom}), while not those of curvature photons.
The curvature luminosity at any given time is calculated by summing all the contribution of all particles in the simulation box,
$L_{\rm cur}=(1/2)\sum_i\alpha(r_i)^2P_{\mathrm{cur}}(\gamma_i)$, where $r_i$ is the position of $i$th particle. 

We apply this model to a BH with $M=10M_\odot$ and $a_*=0.9$.
As for $\Bh$, we implicitly assume that the accretion disk surrounding the magnetosphere is in a highly-magnetized state \citep[MAD;][]{Narayan03, Mckinney12}, likewise in our previous simulations for SMBHs (K20, K22).
MADs are expected to be formed around stellar-mass BHs when the mass accretion rates $\dot{M}$ are much lower than the Eddington rate $\dot{M}_{\rm Edd}\equiv4\pi GMm_{\rm p}/\sT c$ \citep{Cao11, Ioka17, Kimura21, Kimura21b}, and then we can write \citep{Tchekho11}
\begin{equation}
\begin{aligned}
\label{bh}
\Bh&=\dfrac{\phi_{\rm BH}(\dot{M}c)^{1/2}}{2\pi\rg}\\
&\simeq3.6\times10^{7} \left(\dfrac{M}{10M_\odot}\right)^{-1/2}\left(\dfrac{\phi_{\mathrm{BH}}}{50}\right)\left(\dfrac{\dot{M}}{10^{-3}\dot{M}_{\mathrm{Edd}}}\right)^{1/2}\,\rm G.
\end{aligned}
\end{equation}
We set $\Bh=2\pi\times10^7\,\rm G$ in the present work.
The values of the other parameters, $p$, $\epsilon_{min}$, $\tau_0$, and $j_0$ in our simulations are shown in Table~\ref{tab:models}. $\epsilon_{\rm max}$ is set as $10^{-4}$.
We set $\emin=10^{-6}$ as a fiducial value of the peak energy, which corresponds to the optical energy band, consistent with the calculation of photon spectra from stellar-mass BH MADs \citep[c.f.][]{Kimura21}.

The computational domain is set from $r_{\rm min}\simeq1.5\rg$ to $r_{\rm max}\simeq4.3\rg$, so that $r_{\rm min}$ is fairly close to the horizon ($r_\mathrm{H}\simeq1.43r_\mathrm{g}$ for $a_*=0.9$).
We impose the open boundary condition for both boundaries.
The inclination of the simulating region with respect to the BH rotation axis is set as $\theta=30^\circ$.
In every run the number of grid cells exceeds 32000, which was sufficient to resolve the skin depth, and the initial particles per cell number is 45 so that the simulation guarantees convergence (see K20).
A time step is set to satisfy the Courant-Friedrichs-Lewy condition at the inner boundary.

\begin{deluxetable}{ccccc}
\tablenum{1}
\tablecaption{Model parameters of the simulation.}\label{tab:models}
\tablewidth{0pt}
\tablehead{
Model & $j_0$ & $\tau_0$ & $\epsilon_{min}$ & $p$}
\startdata
 LA1 & $-1/2\pi$ & $30$ & $10^{-6}$ & $2.0$ \\
 LA2 & - & $55$ & - & - \\
 LA3 & - & $100$ & - & - \\
 LA4 & - & $175$ & - & - \\
 LA5 & - & $300$ & - & - \\
 LB1 & - & $30$ & - & 2.5 \\
 LB2 & - & $100$ & - & - \\
 LB3 & - & $300$ & - & - \\
 LC1 & - & $30$ & $10^{-5}$ & 2.0 \\
 LC2 & - & $100$ & - & - \\
 LC3 & - & $300$ & - & - \\
 I1 & $-1/2$ & $30$ & $10^{-6}$ & $2.0$ \\
 I2 & - & $100$ & - & - \\
 I3 & - & $300$ & - & - \\
 H1 & $-1$ & $30$ & - & - \\
 H2 & - & $100$ & - & - \\
 H3 & - & $300$ & - & - \\
\enddata
\end{deluxetable}

\begin{figure*}
\hspace*{-2.5cm}
\centering
\begin{minipage}[b]{0.3\linewidth}
 \includegraphics[keepaspectratio, trim = 5 20 100 80, scale=0.42]{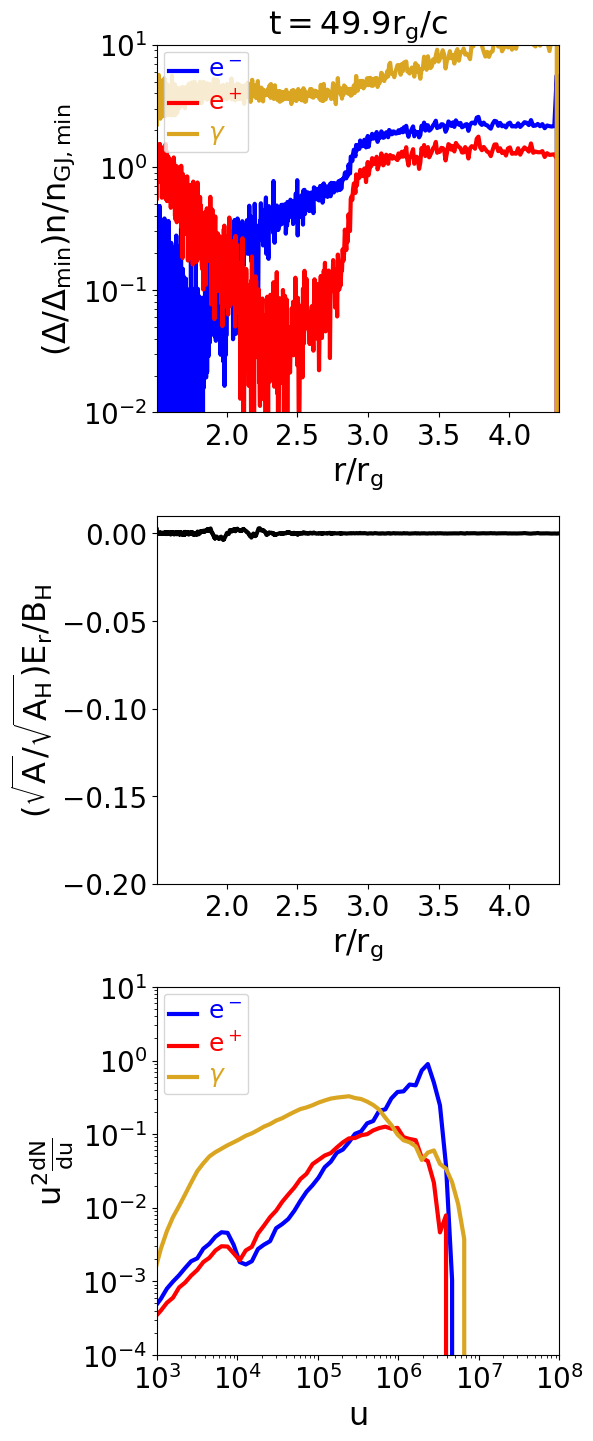}
 \end{minipage}
 \hspace*{0.06\columnwidth}
 \begin{minipage}[b]{0.3\linewidth}
 \includegraphics[keepaspectratio, trim = 5 20 100 80, scale=0.42]{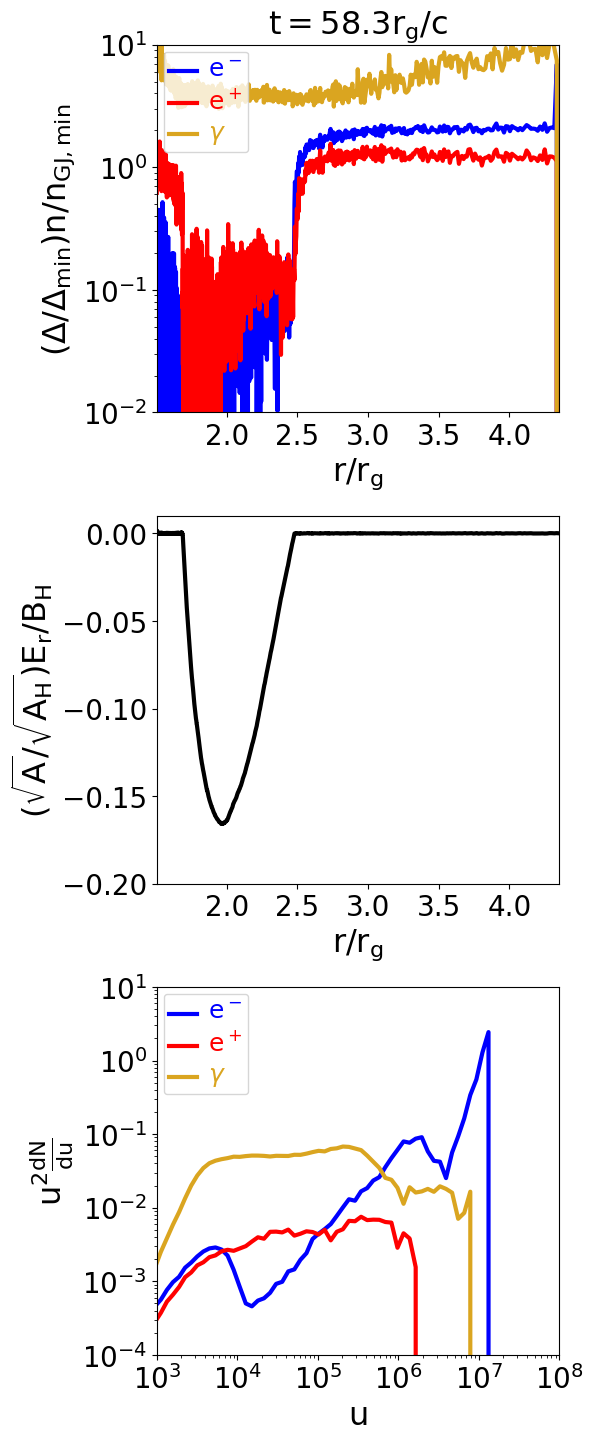}
 \end{minipage}
\hspace*{0.06\columnwidth}
\begin{minipage}[b]{0.3\columnwidth}
\includegraphics[keepaspectratio, trim = 5 20 100 80, scale=0.42]{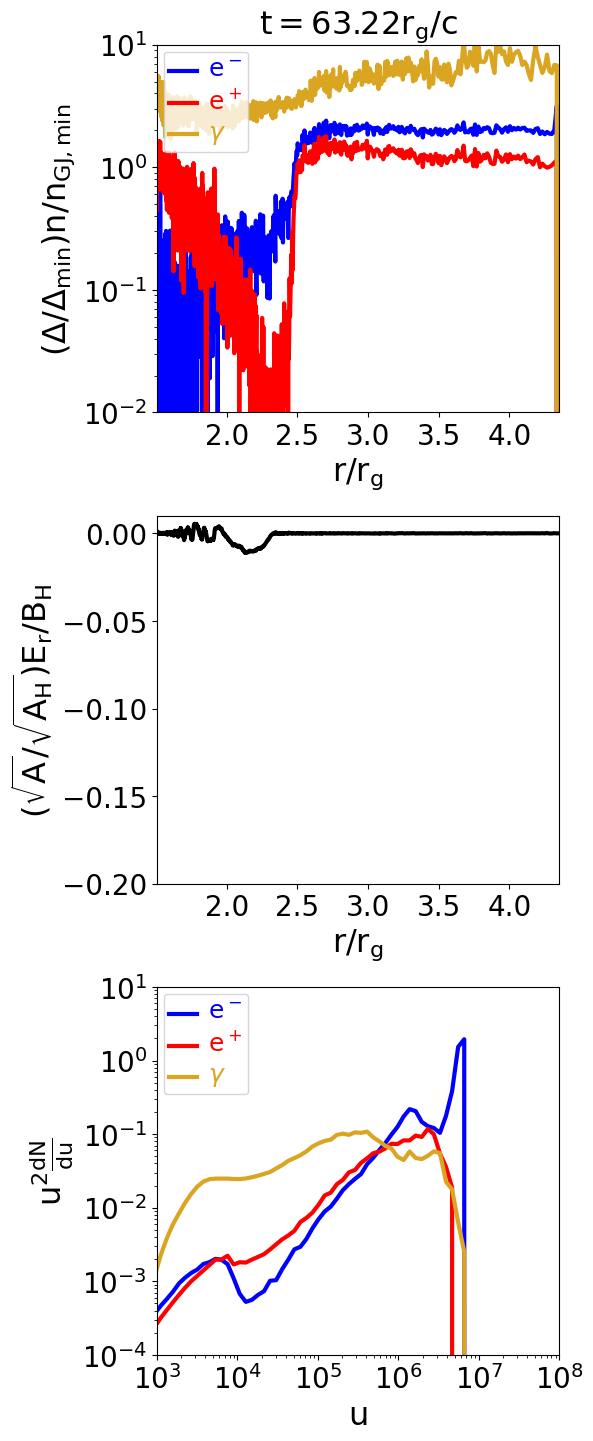}
\end{minipage}
\hspace*{1.5cm}
\caption{Snapshots of the gap at three times in one period of the oscillation in model LA1. The three panels from top to bottom for each of the times are 
(1) the number density of each particle species normalized by the GJ number density at the inner boundary $n_{\rm GJ,min}=\rho_{\rm GJ,min}/e$ ($\Delta_{\rm min}=\Delta(r_{\rm min})$), (2) the electric field distribution normalized by $\Bh$, and (3) the energy distribution of outgoing particles of each species.}
\label{snap}
\end{figure*}

\section{Results}\label{sec:simu_res}
\subsection{Cases of low $|J_0|$
($j_0=-1/2\pi$)}\label{subsec:weak}
We first demonstrate the results for the low $|J_0|$ cases, i.e. $j_0=-1/2\pi$.
Fig.~\ref{snap} shows the evolution of the system for model LA1.
The electric field triggered by inward and outward drifts of plasma develops around the null charge surface ($\rnull\approx2.0$ for $a_*=0.9$), and a quasi-periodic oscillation is observed, similarly as K20, K22. 
Electrons linearly accelerated by the electric field become mono-energetic around $u\sim\gammae\sim10^7$ during the peak time of the electric field ($t=58.3r_\mathrm{g}/c$).

\begin{figure}[t]
\hspace*{0.5cm}
 \includegraphics[keepaspectratio, trim = 5 10 100 10, scale=0.5]{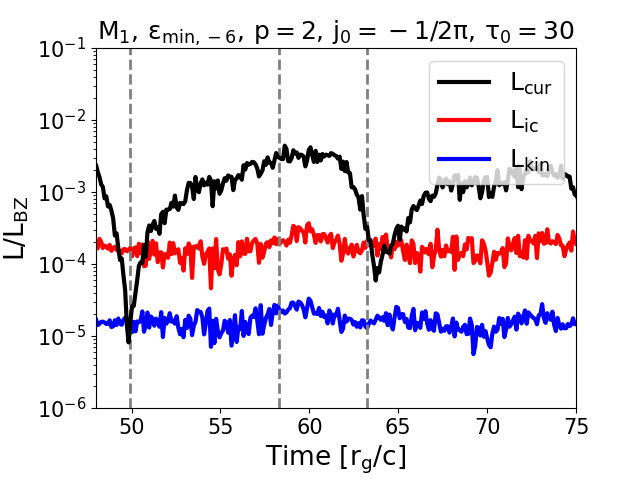}
\caption{Luminosities of curvature emission (\textit{black} line), IC emission (\textit{red} line), and kinetic energy (\textit{blue} line) as ratios to the BZ lumionsity $L_{\rm BZ}$. The vertical dashed lines correspond to the times of the snapshots in Fig.~\ref{snap}.}
\label{lc}
\end{figure}
\begin{figure*}[ht!]
\hspace*{-2.0cm}
\centering
\begin{minipage}[b]{0.45\linewidth}
 \includegraphics[keepaspectratio, trim = 5 80 100 250, scale=0.6]{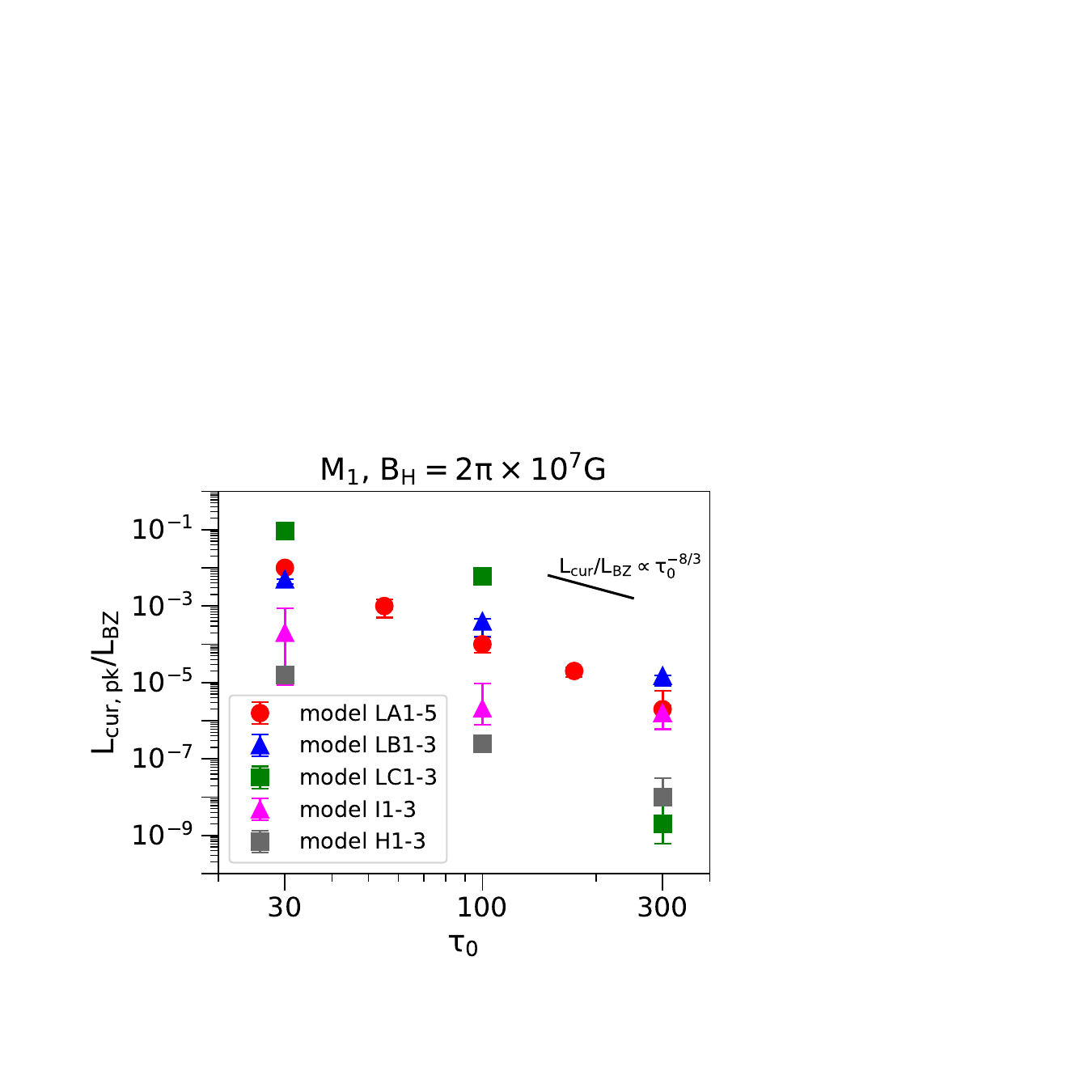}
\end{minipage}
 \hspace*{0.05\columnwidth}
 \begin{minipage}[b]{0.45\linewidth}
 \includegraphics[keepaspectratio, trim = 5 80 100 250, scale=0.6]{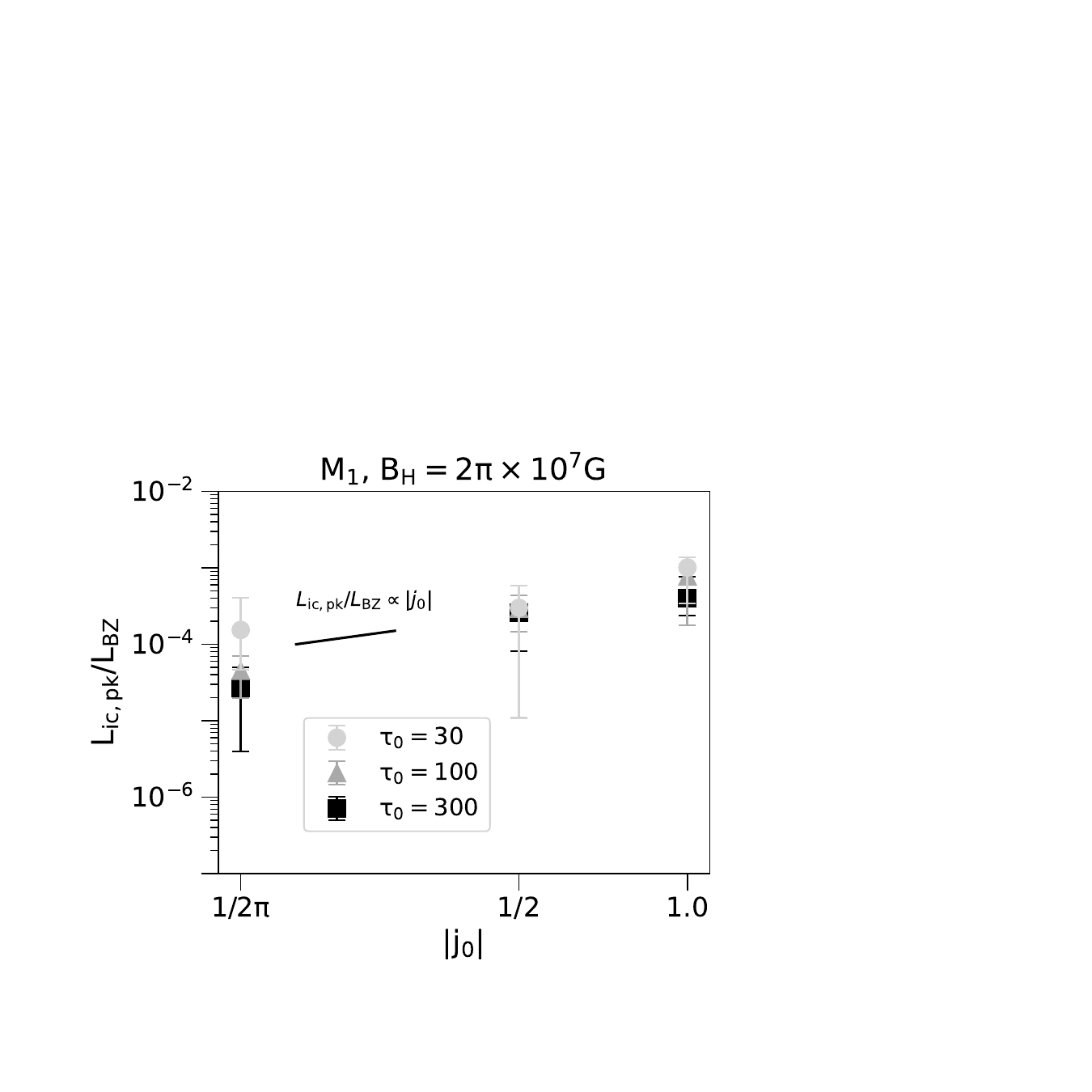}
 \end{minipage}
\caption{
Plots of the peak curvature luminosities normalized by $L_{\rm BZ}$ (\textit{left}) in our simulations with various values of $p, \emin$, and $j_0$ as functions of $\tau_0$ and the peak IC luminosities normalized by $L_{\rm BZ}$ (\textit{right}) in our simulations with $\emin = 10^{-6}$ and $p=2.0$ as functions of $|j_0|$.
}
\label{Ccomp}
\end{figure*}
Fig.~\ref{lc} shows the lumionsities of curvature emission, IC emission, and particle kinetic energies for model LA1.
Here, we normalize them by the BZ luminosity written as
$\Lbz=(\kappa_{\rm B}/4\pi c)\omega_{\rm H}^2\Phi_{\rm H}=(\kappa_{\rm B}\pi c/4)a^2\Bh^2\rg^2$, where $\kappa_{\rm B}$ is set to be 0.053 for the split-monopole magnetic field configuration \citep{Tchekho10}.
We observe luminosity oscillations synchronizing 
with that of the electric field, similarly as the SMBH cases.
The peak values
are $L_{\rm cur,pk}\sim10^{-2}\Lbz$, $L_{\rm ic,pk}\sim10^{-4}\Lbz$, and $L_{\rm kin,pk}\sim10^{-5}\Lbz$.

We then investigate the dependence of gap dynamics on parameters regarding the seed photon properties, 
$\tau_0$, $\epsilon_{\mathrm{min}}$, and $p$.
In the left panel of Fig.~\ref{Ccomp} we show the ratio of peak luminosities of curvature radiation to the BZ luminosity $L_{\mathrm{cur,pk}}/L_{\mathrm{BZ}}$ for models LA-C for various values of $\tau_0$.
$L_{\mathrm{cur,pk}}/L_{\mathrm{BZ}}\propto\tau_0^{\alpha}\,(\alpha\sim-8/3)$ for model LA1-5, which indicates the sensitive dependence of the gap dynamics on the $\gamma\gamma$ pair production efficiency, is similar to that shown in K22.
We also find that the effect of changing the spectral hardness $p$ on the gap behavior is minor.
This is because most of the IC scatterings occur in Klein-Nishina (KN) regime, for which the $\gamma\gamma$ pair production optical depth is independent on $p$ (see K20).
Change in the peak energy $\emin$, however, strongly affects the dynamics similarly to K20, K22.
For $\epsilon_{\rm min} = 10^{-5}$, the width and period of the gaps are $\sim10$ times larger 
than those for $\epsilon_{\rm min} = 10^{-6}$ in the cases of $\tau_0=30$ and $100$ (compare models LC1 and LC2 with models LA1 and LA3, respectively).\footnote{Due to much longer gap oscillation periods for model LC1 and LC2 than those for the other models, we observe only a few oscillations for LC1 and 2 in our simulations, which might affect the analysis.}
In contrast, a very narrow and highly intermittent gap appears for $\epsilon_{\rm min} = 10^{-5}$ in the case of $\tau_0=300$ (model LC3).

\subsection{Cases of intermediate and high $|J_0|$}
\label{subsec:j0p5}
We next perform the simulations for $j_0 = -1/2$ and $-1$. 
In these cases, the gap does not appear to be clearly periodic and the electric field strength is weaker than the cases of $j_0=-1/2\pi$ with the same $\tau_0$, $p$, and $\emin$. This is because the higher electric current $|J_0|$ leads to the higher density of plasma flowing in the gap, which makes it easier to screen the electric field.

In Fig.~\ref{Ccomp} we compare $L_{\rm cur,pk}$ and $L_{\rm ic,pk}$ for different values of $j_0$. 
$L_{\rm cur,pk}$ for $j_0=-1/2$ is $10^{-1}-10^{-2}$ times smaller than that for $j_0=-1/2\pi$ for all of the $\tau_0$ values. This results from the weaker electric field and lower efficiency of particle acceleration. 
$L_{\rm ic,pk}$, on the contrary, is slightly larger
for lower $j_0$, due to an increase of the plasma number density in the gap.
Its scaling is roughly $L_{\rm ic,pk}\propto |j_0|$.

The dependence of the particle number density on $j_0$ at the peak time of electric field is $\left.(\Delta/\Delta_{\rm min})n/n_{\rm GJ,min}\right|_{ r=r_{\rm null}}\sim0.08$ for $j_0=-1/2\pi$, $\sim0.3$ for $j_0=-1/2$, and $\sim0.5$ for $j_0=-1$, where $n_{\rm GJ,min}=\rho_{\rm GJ}(r_{\rm min})/e$ and $\Delta_{\rm min}=\Delta(r_{\rm min})$. Note that the density does not appear to be dependent on $\tau_0$, $p$, and $\epsilon_{\rm min}$.
This $j_0$ dependence can be understood analytically, by using the Ampere's law (Eq.~\ref{rAmp}): 
The development of $E_r$ stops at the peak time, then we have
\begin{equation}
\label{amppk}
\Sigma j^r (r_{\rm null})\approx J_0.
\end{equation}
Since the positrons (electrons) carrying $j^r$ are accelerated to relativistic velocity $v_i\approx-1(+1)$ around the null surface, 
$j^r (r_{\rm null})$ can be written by using the relation in the number density of electrons and positrons, $n_{\rm +}(r_{\rm null})\approx n_{\rm -}(r_{\rm null}) = n(\rnull)$
\begin{equation}
\begin{aligned}
\label{jr}
j^r(r_{\rm null})=-\dfrac{\Delta}{\sqrt{A}}2en({\rnull}).
\end{aligned}
\end{equation}
Then, we obtain
\begin{equation}
\label{nnullest}
\dfrac{\Delta(\rnull)}{\Delta_{\rm min}}\dfrac{n(\rnull)}{n_{\rm GJ,min}}\approx-0.58j_0,
\end{equation}
where we use Eqs.~(\ref{rhoGJ}) and (\ref{J0}). Eq.~(\ref{nnullest}) is consistent with our findings in the simulations. 
We utilize this relation in our semi-analytic model described in the next section.

\section{Semi-Analytic Model of the Gap for Low Current Cases} \label{sec:semiana}
K22 built a semi-analytic model for estimate of peak curvature luminosities from the gaps for SMBH cases with $j_0=-1/2\pi$. 
In Section~\ref{sec:simu_res}, we have seen the overall similarities of the gap dynamics for stellar-mass BH cases to SMBH cases, which motivate us to construct a semi-analytic model applicable to stellar-mass BH cases. 
From the model of K22, we improve the model by taking into account GR effects and the exact expression of KN cross section. This model helps clarify the dynamics of gaps in BH magnetospheres and predict the gamma-ray luminosities and spectra from the gaps over broad ranges of masses $M$ and accretion rates $\dot{M}$.

We assume the electric field in the gap to be steady, since the oscillation periods in many cases are considerably longer than the light crossing of the gaps.
The nearly mono-energetic distribution of outgoing electrons around the peak time of electric field (see Fig.~\ref{snap}) allows us to consider that all of the electrons would experience similar acceleration and radiative interactions. We solve the equation of motion and the scattered photon propagation for one representative electron, and determine the solutions of gap inner and outer boundaries $\rin$ and $\ro$ for which sufficient electron positron pairs are created inside the gap.
The corresponding maximum Lorentz factor $\gamma_{\rm e,max}$, and the gamma-ray peak luminosity $L_{\rm cur,pk}$ are also calculated.
Differences between our model and the previous analytic steady gap models \citep[e.g.,][]{Hirotani18} will be briefly discussed in Section~\ref{sec:dandc}.

\subsection{Modeling the Dynamics} \label{subsec:dynam}
\subsubsection{Equation of motion in the gap} \label{ssub:eom}
We first solve the equation of motion of an outgoing electron.
We define $r_{\rm in}$ as the initial radius.
Our 1D simulations assume $u_\phi = 0$ for all the particles, and the electron is rapidly accelerated to a highly relativistic velocity. Then we have $\alpha dt\approx\sqrt{g_{\mathrm{rr}}} dr$ for the electron's motion, so that Eq.~(\ref{EoM}) can be rewritten as
\begin{equation}
\label{en}
\dfrac{d\gammae}{dr}=-\dfrac{\sqrt{g_{\mathrm{rr}}}}{\me c^3}\left[eE_r(r,\,r_{\mathrm{in}},\,B_{\rm H})c+P_{\mathrm{cur}}(\gammae)+P_{\mathrm{ic}}(\gammae)\right],
\end{equation}
where we have added the IC cooling term $P_{\rm ic}$. (In the simulations IC cooling is treated via the Monte-Carlo approach.)
We confirmed that the gravitational force (1st term of the rhs of Eq.~\ref{EoM}) is negligible for the energy range achieved in our simulations, $\gammae \lesssim 10^7$. 

The simulations show for the number densities of electrons and positrons $n_+\approx n_-\ll n_{\rm GJ,min}$ around the peak time of gap electric field.
This enables us to approximate the electric field by the vacuum ($j^t=0$) solution of Gauss' equation (Eq.~\ref{rGauss})
\begin{equation}
\label{Efield}
E_{\mathrm{r}}(r,\,r_{\mathrm{in}},\,B_{\rm H})=-\frac{4\pi}{\sqrt{A}}\int^{r}_{r_{\mathrm{in}}}dr'\,\Sigma\rho_{\mathrm{GJ}}(r',\,B_{\rm H}),
\end{equation}
where we have imposed the boundary condition $E_{\rm r}(r_{\rm in})=0$.

The IC emitting power is written as
\begin{equation}
\label{ic}
P_{\mathrm{ic}}(\gammae,\,\tau_0)=\sigma_{\mathrm{ic}}n_\mathrm{s}(\epsilon_{\mathrm{min}})\epsilon_{\mathrm{ic}}\me c^3=\dfrac{\tau_0}{r_{\rm g}}\dfrac{\sigma_{\rm ic}}{\sigma_{\rm T}}\epsilon_{\mathrm{ic}}m_\mathrm{e}c^3.
\end{equation}
Here $\sigma_{\rm ic}$ is the KN cross section
\begin{equation}
\label{sic}
\sigma_{\mathrm{ic}}=\frac{3}{8}\sigma_\mathrm{T}\dfrac{1}{x^2}\left[4+\dfrac{2x^2\left(1+x\right)}{\left(1+2x\right)^2}+\dfrac{x^2-2x-2}{x}\log(1+2x)\right],
\end{equation}
where $x=\gammae\epsilon_{\mathrm{min}}$.
$\epsilon_{\mathrm{ic}}$ is the energy of the scattered photon approximately given by
\begin{equation}
\label{eic}
\epsilon_{\mathrm{ic}}=0.3\times\left\{
\begin{aligned}
&\gammae^2\epsilon_{\mathrm{min}}\hspace{10pt}(\gammae\epsilon_{\mathrm{min}}\lesssim1)\\
&\gammae\hspace{30pt}(\gamma_\mathrm{e}\epsilon_{\mathrm{min}}\gtrsim1)
\end{aligned}
\right..
\end{equation}

The Lorentz factor of an electron at the $k$th step $r_k$ of the integration of Eq.~(\ref{en}) is determined as  $\gamma_{\mathrm {e},k}\equiv\gamma_{{\rm e}}(r_k)=\min\{\gamma_{\mathrm{e,acc},k}, \,\gamma_{\mathrm{e,cur},k},\,\gamma_{\mathrm{e,ic},k}\}$ (see Fig.~\ref{model}), where $\gamma_{\mathrm{e,acc},k}$ is the value for the acceleration-dominant regime,
\begin{equation}
\label{gac}
\gamma_{\mathrm{e,acc},k}=\gamma_{\mathrm{e},k-1}-\sqrt{g_{\mathrm{rr}}}\dfrac{eE_\mathrm{r}}{\me c}dr.
\end{equation}
$\gamma_{\mathrm{e,cur},k}$ and $\gamma_{\mathrm{e,ic},k}$ are the values determined by the balances of the acceleration with the curvature and IC coolings, respectively,
\begin{equation}
\label{gc}
-eE_\mathrm{r}(\,r_{k-1},r_{\mathrm{in}},\,B_{\rm H})c=\left\{\begin{aligned}
&P_{\mathrm{cur}}(\gamma_{\mathrm{e},\mathrm{cur},k})\\
&P_{\mathrm{ic}}(\gamma_{\mathrm{e},\mathrm{ic},k},\tau_0)
\end{aligned}\right..
\end{equation}

\subsubsection{$\gamma\gamma$ Pair production} \label{ssub:pair}
Next, we compute $\gamma\gamma$ pair productions in the gap.
The number of photons scattered by the electron as it propagates from $r_{k-1}$ to $r_{k}$ is calculated by considering that seed photons with $\epsilon_{\rm min}$ are most abundant,
\begin{equation}
\delta N_{\mathrm{\gamma}, k}=\dfrac{dr}{l_{\mathrm{ic}}(\epsilon_{\mathrm{min}},\gamma_{{\rm e},k},\tau_0)},
\end{equation}
where $l_{\mathrm{ic}}$ is the mean free path of IC scattering,
\begin{equation}
\begin{aligned}
\label{lkn}
&l_{\mathrm{ic}}(\epsilon_{\mathrm{min}}, \gammae,\tau_0)=\dfrac{1}{\sqrt{g_{\rm rr}}}\dfrac{1}{n_\mathrm{s}(\epsilon_{\mathrm{min}})\sigma_{\mathrm{ic}}}
=\dfrac{\rg}{\sqrt{g_{\mathrm{rr}}}\tau_0}\dfrac{\sT}{\sigma_{\rm ic}}.
\end{aligned}
\end{equation}

We consider that all the scattered photons annihilate with the seed photons after propagating about the mean free path for the $\gamma\gamma$ pair annihilation $l_{\gamma\gamma}$,
\begin{equation}
\begin{aligned}
\label{lgg}
l_{\gamma\gamma}(\epsilon_{\mathrm{min}},\gamma_\mathrm{e},\tau_0)&=\dfrac{1}{\sqrt{g_{\mathrm{rr}}}}\frac{1}{n_\mathrm{s}(\epsilon_2)\sigma_{\gamma\gamma}}\\
&=\dfrac{10\,\epsilon_{\mathrm{ic}}\epsilon_2r_\mathrm{g}}{\sqrt{g_{\rm rr}}\tau_0}\left(\frac{\epsilon_2}{\epsilon_{\mathrm{min}}}\right)^{p},
\end{aligned}
\end{equation}
where $\epsilon_2$ is the energy of the annihilating seed photon, defined as $\epsilon_2=\max\{\epsilon_{\mathrm{min}},\,\epsilon_{\mathrm{ic}}^{-1}\}$, and we use $\sigma_{\gamma\gamma}\sim0.1\sigma_\mathrm{T}(\epsilon_{\mathrm{ic}}\epsilon_2)^{-1}$ for the $\gamma\gamma$ pair annihilation cross section.
Thus, the number of pairs created by photons scattered between the $k-1$th and the $k$th step is
\begin{equation}
\delta N_{\mathrm{pair},k}(r_{\mathrm{ppf},k},\gamma_{\mathrm{e},k})=\delta N_{\gamma,k},
\end{equation}
where the radius of the `pair production front' $r_{\mathrm{ppf}}$ of the scattered photons is given by
\begin{equation}
\label{rpcf}
r_{\mathrm{ppf},k}\equiv r_{\mathrm{ppf}}(r_k,\gamma_{\mathrm{e},k})=r_k+l_{\gamma\gamma}(\epsilon_{\mathrm{min}},\,\gamma_{\mathrm{e},k}).
\end{equation}

The total number of newly created pairs from $r_{\mathrm{in}}$ to the position $r$ is calculated as 
\begin{equation}
\label{pairden}
N_{\mathrm{pair}}(r)=\sum_{r_{\mathrm{ppf},k}\leq r}\delta N_{\mathrm{pair},k}(r_{\mathrm{ppf},k}).
\end{equation}
\begin{figure}
\centering
\includegraphics[keepaspectratio, scale=0.6, trim = 100 50 100 110]{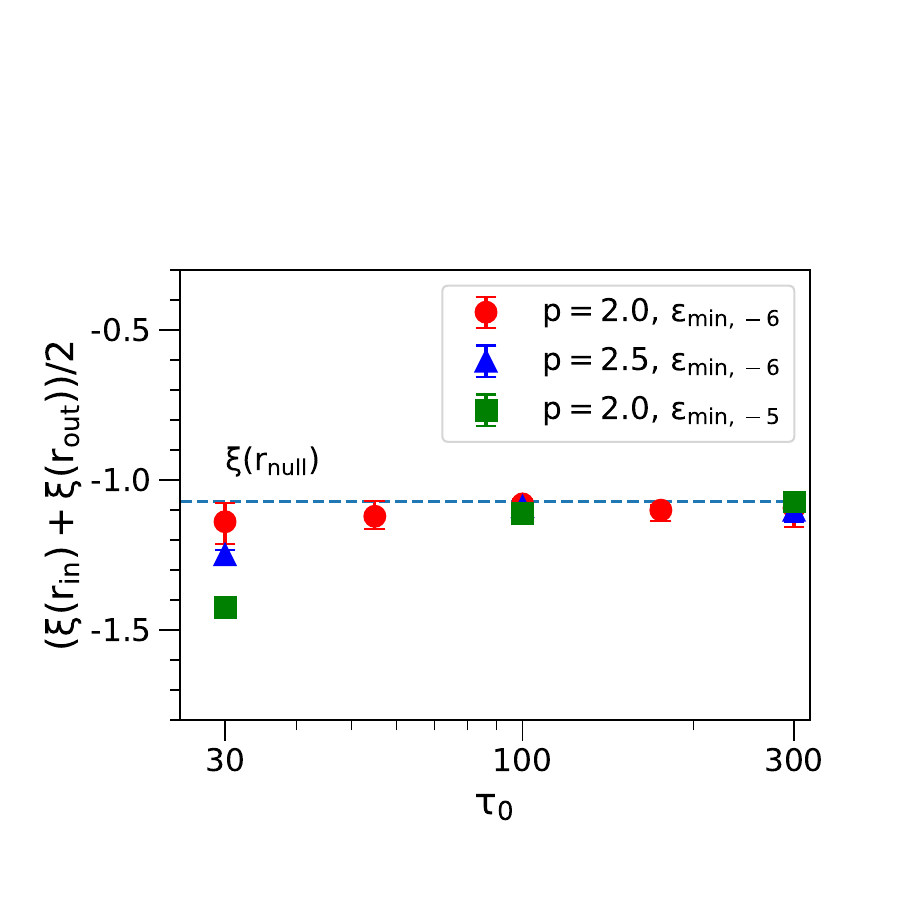}
\caption{Middle points of the gaps in the $\xi$ coordinate from the simulation results. The horizontal dashed line represents the position of the null charge surface.}
\label{gboundary}
\end{figure}
\begin{figure*}
\centering
\hspace*{-1cm}
\includegraphics[keepaspectratio, trim = 100 10 100 20, scale=0.6]{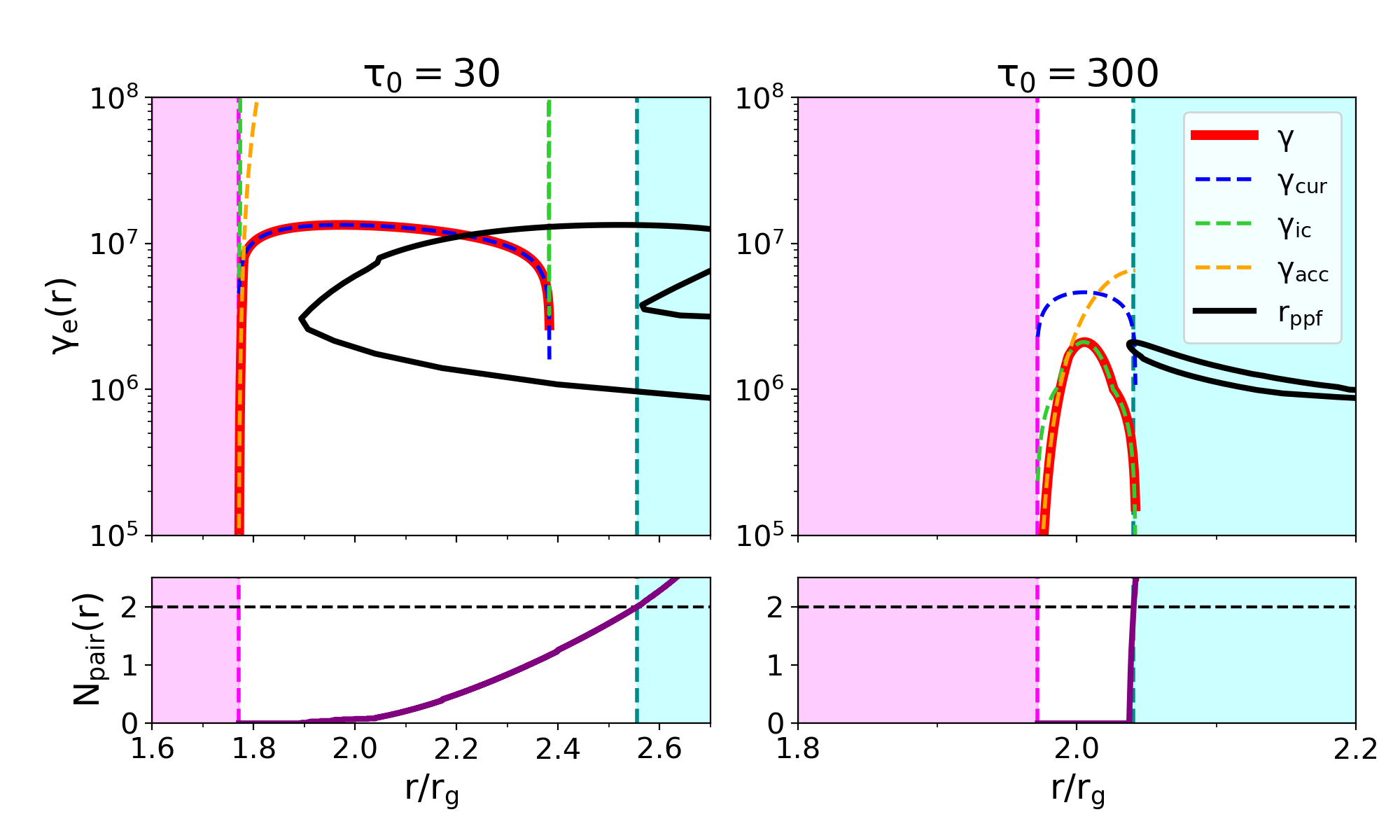}
\caption{Results of our semi-analytic model for different $\tau_0$ (\textit{left}: $\tau_0=30$, \textit{right}: $\tau_0=300$), with $\epsilon_{\rm min}=10^{-6}, p=2.0,$ and $B_{\rm H}=2\pi\times10^7\,\mathrm{G}$. The top panels show the evolutions of $\gamma_{\mathrm{e,acc}}$ (\textit{orange} dashed line), $\gamma_{\mathrm{e,cur}}$ (\textit{blue} dashed line), and $\gamma_{\mathrm{e,ic}}$ (\textit{light-green} dashed line). 
The \textit{red} solid curve represents the solution of $\gamma_\mathrm{e}$. 
The \textit{black} solid line represents the radii at which the scattered photons annihilate with the seed photons, $r_{\rm ppf}$. 
The bottom panel show the evolutions of number of pairs $N_{\rm pair}$.
The vertical dashed lines indicate $r_{\rm in}$ (where $N_{\rm pair} = 0$) and $r_{\rm out}$ (where $N_{\rm pair} = 2.0$), and the white regions correspond to the gaps.}

\label{model}
\end{figure*}
\subsubsection{Gap width and luminosity} \label{ssub:pair}
We iteratively calculate the total amount of created pairs for trial values of $r_{\rm in}$ with $\rh < r_{\rm in} < r_{\rm null}$, until we find the outer boundary of the gap $r_{\mathrm{out}}$ that satisfies the following two criteria:
(1) The total number of newly created pairs between $r_{\rm in}$ and $r_{\rm out}$ is
\begin{equation}
\label{ncri}
N_{\mathrm{pair}}(r_{\mathrm{out}})=2.0.
\end{equation}
It is expected that this number should be order of unity to have the electric field stop growing. We found that with the value of $2.0$ our semi-analytic model well fits the simulation results.
(2) The gap is symmetric in the $\xi$ coordinate (see Appendix~\ref{space}), i.e.,
\begin{equation}
\label{rcri}
\dfrac{\xi(r_{\mathrm{out}})+\xi(r_{\mathrm{in}})}{2}=\xi(r_{\mathrm{null}}),
\end{equation}
which is satisfied in the simulation results, as shown in Fig.~\ref{gboundary}.

For the solution of $r_{\rm in}$ and $r_{\rm out}$ obtained above, we calculate the peak curvature luminosity by
\begin{equation}
\begin{aligned}
\label{Lcur}
L_{\mathrm{cur}}&=2\pi\int^{r_{\mathrm{out}}}_{r_{\mathrm{in}}}\alpha^2(r)P_{\mathrm{cur}}\left(\gamma_\mathrm{e}(r)\right)n(\rnull)\Sigma(r)dr\\
&\approx0.39\frac{e^2c}{r_\mathrm{g}^2}\dfrac{\Delta_{\rm min}}{\Delta(\rnull)}n_{\rm GJ,min}\int^{r_{\mathrm{out}}}_{r_{\mathrm{in}}}\alpha^2\gamma_e^4\Sigma dr,
\end{aligned}
\end{equation}
where we have assumed that the number density of emitting particles is constant throughout the gap, and evaluate it by $n(\rnull)$ (see Eq.~\ref{nnullest}).

\subsection{Examples of Model Solutions}\label{subsec:ex}
Fig.~\ref{model} shows 
solutions of our semi-analytic model for $\tau_0 = 30$ and $300$, with the other parameters $\epsilon_{\rm min}=10^{-6}$, $p=2$, and $B_{\rm H}=2\pi\times10^7\,\mathrm{G}$ (same as models LA).
The top panels show the energy evolutions of the electron.
The vertical dashed lines 
indicate $r_{\mathrm{in}}$ (\textit{magenta}) and $r_{\mathrm{out}}$ (\textit{cyan}) determined as the solutions, and the white regions correspond to the gaps.
The dashed curves show the evolutions of Lorentz factors calculated by Eqs.~(\ref{gac}) and (\ref{gc}) for the solutions.
Determined $\gamma_{\mathrm{e}}$ at each $r$ is shown by the \textit{red} solid line.
We also show $r_{\rm ppf}$ for each $r$ and $\gammae$.

As can be seen in Fig.~\ref{model},
the maximum electron Lorentz factors $\gamma_{{\rm e, max}} \sim 10^7$ for $\tau_0 = 30$ and $\gamma_{{\rm e,max}} \sim 2 \times 10^6$ for $\tau_0 = 300$ are determined by the curvature cooling and the IC cooling, respectively.
This results from the different particle energy dependence of the two cooling mechanisms as well as the $\tau_0$ dependence of the IC scattering mean free path:
For $\tau_0 = 30$, $l_{\rm ic}$ becomes larger compared with the case of $\tau_0 = 300$ which leads to lower $\delta N_{\rm pair}$.
Thus, the solution with larger gap width is obtained, in which the maximum Lorentz factor reaches $\gammae \sim 10^7$.
In this regime the curvature cooling is more efficient than the IC cooling because $P_{\mathrm{cur}}\propto\gamma_\mathrm{e}^4$ and the IC scattering is deeply in the KN regime.
On the other hand, for $\tau_0 = 300$, $\gamma_{{\rm e,max}}$ is regulated by the IC cooling in the Thomson regime or marginally in the KN regime.
This transition of dominant cooling process matches with the trend of the gap dynamics and energetics in the simulations, in which we see that the gaps become narrower for higher $\tau_0$ and $L_{\rm cur,pk} \lesssim L_{\rm ic,pk} \sim 10^{-4} \Lbz$ for $\tau_0 \gtrsim 100$ (see Fig.~\ref{Ccomp}).
\begin{figure*}[ht!]
\hspace*{-0.5cm}
\centering
\begin{minipage}[b]{0.45\linewidth}
 \includegraphics[keepaspectratio, trim = 5 30 100 95, scale=0.6]{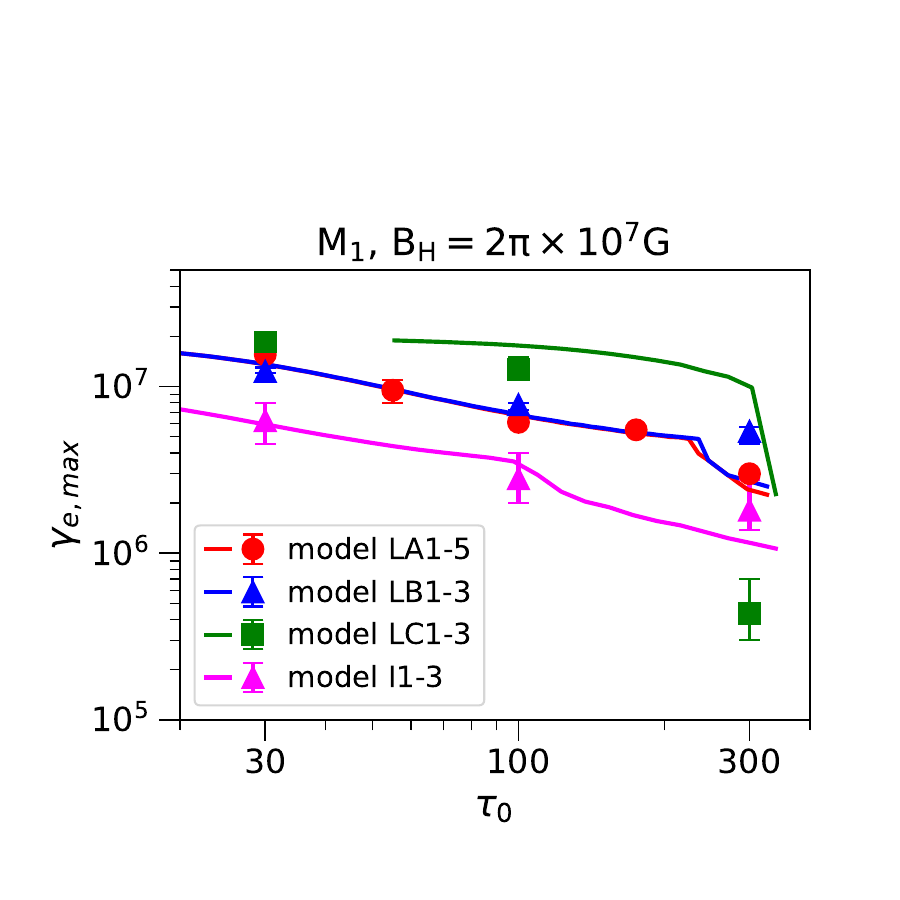}
 \end{minipage}
 \hspace*{0.12\columnwidth}
 \begin{minipage}[b]{0.45\linewidth}
 \includegraphics[keepaspectratio, trim = 5 30 100 95, scale=0.6]{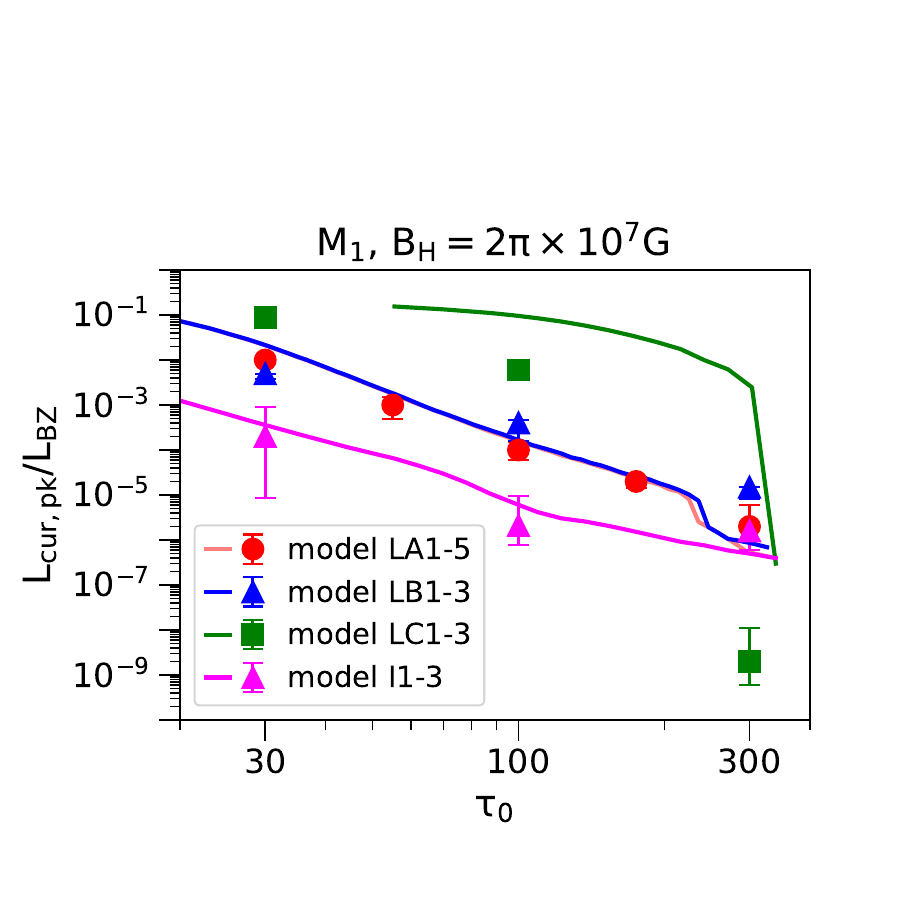}
 \end{minipage}
\caption{
Comparison between the simulation results (\textit{points}) and the semi-analytic model solutions (\textit{lines}). We plot $\gamma_{\rm e,max}$ (\textit{left}) and $\Lcur/\Lbz$ (\textit{right}) as functions of $\tau_0$.
}
\label{comp}
\end{figure*}

\subsection{Model-Simulation Comparison}\label{subsec:comp}
We compare the simulation results and semi-analytic model solutions of $\gamma_{\mathrm{e,max}}$ and $L_{\mathrm{cur,pk}}/L_{\mathrm{BZ}}$ for $j_0=-1/2\pi$ in Fig.~\ref{comp}.
For models LA and LB, the semi-analytic solutions
show that $\gamma_{\rm e,max}$ and $L_{\mathrm{cur,pk}}/L_{\mathrm{BZ}}$ are decreasing functions of $\tau_0$ with a break around $\tau_0\sim100-200$, and well fit the simulation results particularly for $\tau_0 < 200$.
For models LC2 and LC3 we obtain $\sim10^2$ times higher $L_{\rm cur,pk}/\Lbz$ than simulation results, and for model LC1 we could not find a solution, but the overall dependence of $\gamma_{\rm e,max}$ and $L_{\rm cur,pk}/\Lbz$ on $\tau_0$ are similar to what we observe in the simulations.

The breaks seen at $\tau_0\gtrsim100$ in all the solutions are due to the transition from the curvature-dominant state to the IC-dominant state, as explained in \ref{subsec:ex}.
The luminosity drop along with this transition is prominent for model LC1-3, which is consistent with the simulation results.
We can conclude that the energy and $\tau_0$ dependences of the two cooling rates affect the dynamics and energetics of the gaps.

\subsection{Model for $j_0=-1/2$ Cases}\label{jp5model}
In the simulations for $j_0 = -1/2$, the time evolution of the gap does not exhibit clear periodicity.
Also, we observe $n_+\neq n_-$ around the peak time except at the middle point of the gap, hence the assumption of the vacuum electric field is not necessarily valid.
Nevertheless, we can build an effective model consistent with the average values of $\gamma_{\rm e,max}$ and $L_{\rm cur,pk}/\Lbz$ in the simulation results by adding some modifications to the model for $j_0 = -1/2\pi$.
We use $1/10$ times weaker $\Er$ than that calculated by Eq.~(\ref{Efield}) and use a different condition on $N_{\rm pair}$ instead of Eq.~(\ref{ncri}),
\begin{equation}
\label{ncri2}
N_{\mathrm{pair}}(r_{\mathrm{out}})=\dfrac{1}{\pi|j_0|}.
\end{equation}
Note that the condition of Eq.~(\ref{ncri}) for $j_0 = -1/2\pi$ is consistent with Eq.~(\ref{ncri2}).
We present the model solutions for $j_0=-1/2$ (model I1-3) in Fig.~\ref{comp}.
As seen, the solutions well fit the simulation results.

\section{Gamma-Ray Signals from Isolated Stellar-Mass BHs}\label{sec:impli}
In this section, we estimate the gamma-ray luminosities from magnetospheres of stellar-mass BHs in the Galaxy based on our semi-analytic model.
About $10^9$ stellar-mass BHs are thought to exist in the Galaxy \citep[e.g.][]{Shapiro86, Sartore10, Caputo17, Abrams20}.
Many of those are isolated stellar-mass BHs (IBHs), wandering in the interstellar space \citep[e.g.][]{Olejak20}.
However, we have not detected any hints of electromagnetic signatures related to them, except for a microlensing event \citep[OGLE-2011-BLG-0462/MOA-2011-BLG-191,][]{Sahu22, Lam22, Prze22, Lam23}.
Recent studies have proposed a possibility of ISM accretion onto IBHs with an efficient accumulation of the magnetic flux \citep{Cao11, Barkov12, Ioka17, Kimura21}.
The magnetic field at the BH vicinity in the case of rapidly-spinning IBHs could be strong enough for their accretion disks to become MAD states \citep[e.g.][]{Kaaz23}.
Then, the gamma-ray emission from gaps in their magnetospheres would be detectable \citep{Hirotani16_2,Song17,Hirotani18}.
\begin{figure*}[ht!]
\hspace*{2.25cm}
\centering
\begin{minipage}[b]{0.4\linewidth}
 \includegraphics[keepaspectratio, scale=0.6, trim = 50 100 200 250]{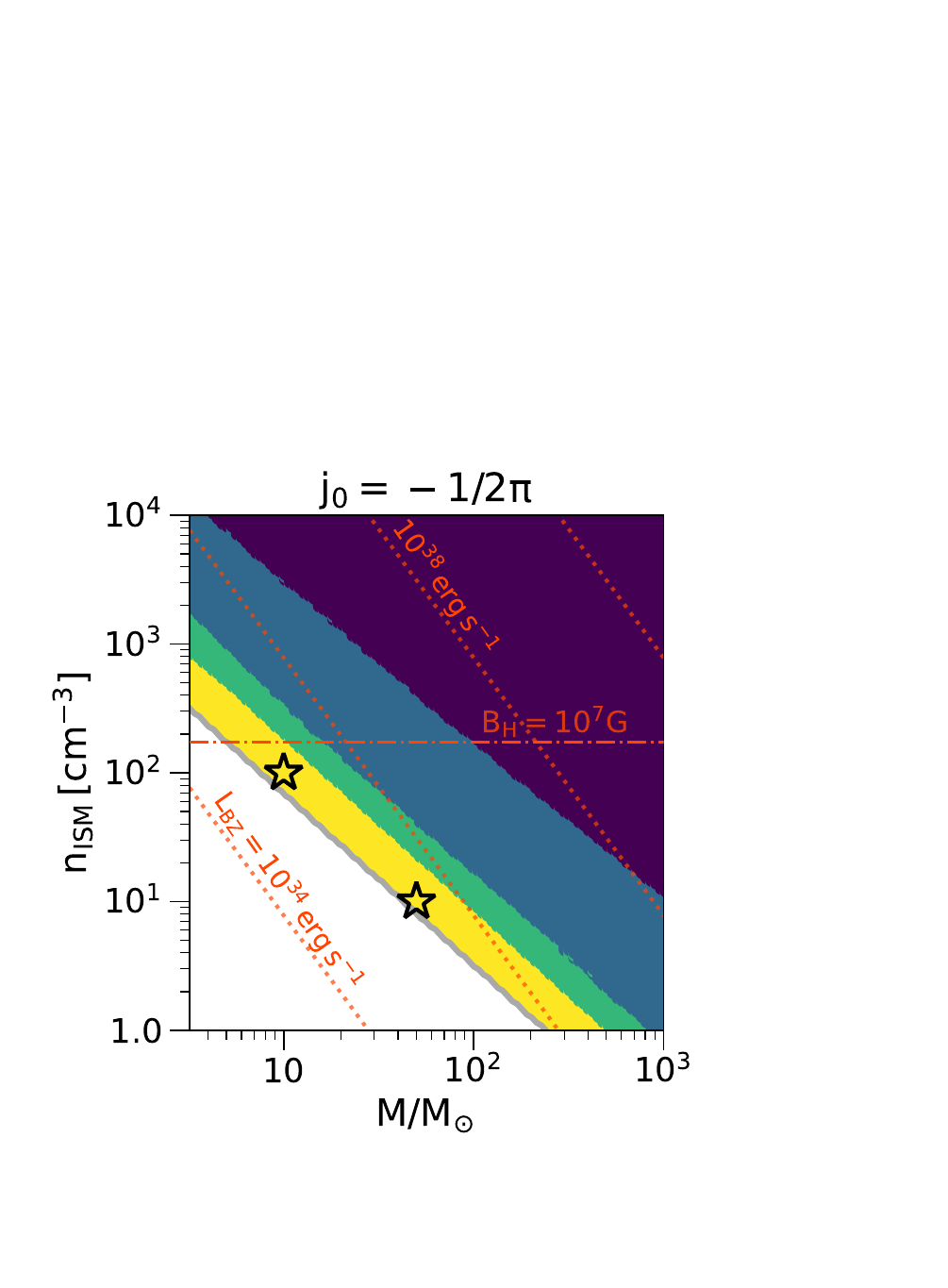}
 \end{minipage}
 \hspace*{0.0\columnwidth}
 \begin{minipage}[b]{0.4\linewidth}
 \includegraphics[keepaspectratio, scale=0.6, trim = 150 100 100 250]{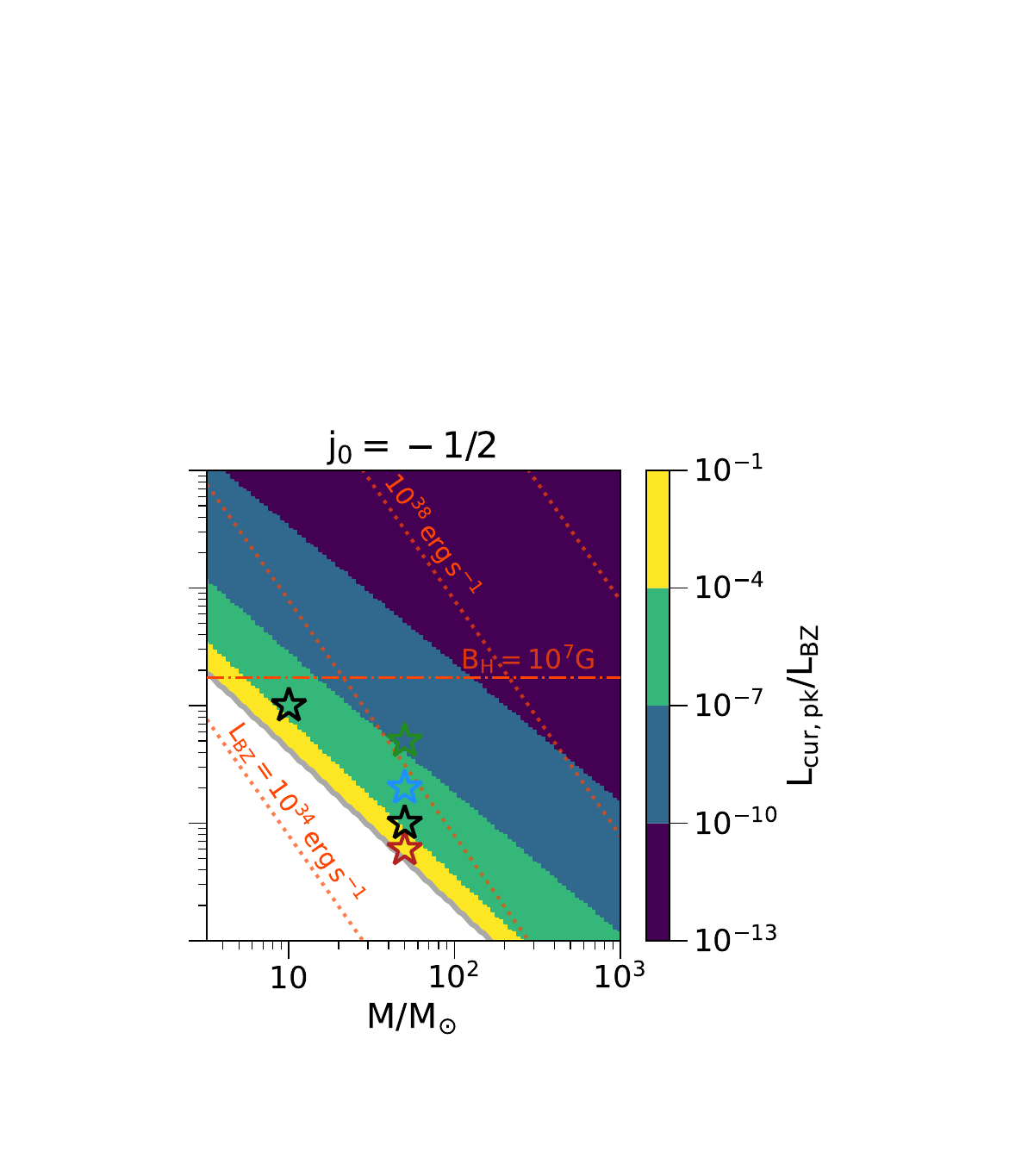}
 \end{minipage}
 \hspace*{2.0cm}
\caption{The color maps of $L_{\rm cur,pk}/\Lbz$ as functions of $M$ and $\nism$ for $j_0=-1/2\pi$ (\textit{left}) and $j_0 = -1/2$ ({\textit{right}}). The white regions indicate the parameter ranges that have no semi-analytic solutions. The stars indicate the parameter sets which we use in the spectral calculations for Fig.~\ref{spectrum} and \ref{spec_ncomp}. We also plot contours of $\Lbz$ with \textit{dotted} lines and $\Bh=10^7\rm G$ with \textit{dot-dashed} line.}
\label{Mnp}
\end{figure*}

When evaluating the input parameters of the semi-analytic model, $\emin$, $\tau_0$, and $\Bh$, we use the one-zone MAD model of \citet{Kimura21}, in which $\dot{M}$ is assumed to be the Bondi-Hoyle-Lyttleton accretion rate.
We consider that thermal synchrotron emission from the MADs as the seed photons for IC scatterings and $\gamma\gamma$ pair productions in the gap.
The magnetic field strength in the MADs $B$ is derived as a function of $M$ and the density of surrounding ISM $\nism$ for the fixed viscosity parameter $\alpha$ and plasma beta $\beta$.
The electron temperature $T_{\rm  e}$ is also determined as a function of $M$ and $\nism$, by equating the electron heating rate in the MAD and the cooling rate via synchrotron radiation.
Then, we can derive $\emin$, $\tau_0$ as
\begin{equation}
\label{emincal}
\begin{aligned}
\emin&=x_{\rm M}\dfrac{3}{4\pi}\dfrac{heB}{\me^2 c^3}\theta_{\rm e}^2\\
&\simeq3.8\times10^{-6}\left(\dfrac{x_{\rm M}}{25}\right)M_1^{-1}n_{\rm ISM,2}^{-1/2}\mathcal{R}_1^{-1/4}\alpha_{-0.5}^{3/2}\beta_{-1}^{1/2},
\end{aligned}
\end{equation}
and
\begin{equation}
\begin{aligned}
\label{t0cal}
\tau_0&=\dfrac{4\pi}{c}\dfrac{\nu_{\rm syn,pk}L_{\nu_{\rm syn,pk}}}{4\pi^2R^2\emin\me c^2}\sT\rg\\
&\simeq39M_1^{2}n_{\rm ISM,2}^{3/2}\mathcal{R}_1^{-7/4}\alpha_{-0.5}^{-3/2}\beta_{-1}^{-1/2},
\end{aligned}
\end{equation}
where $\theta_{\rm e}=\kb T_{\rm e}/\me c^2$ denotes the normalized electron temperature,
$R=\mathcal{R}\rg$ the size of MAD,
and
$x_{\rm M}$ a numerical factor ($x_{\rm M}\sim25$ for the synchrotron-self-absorption thin limit).
$\nu_{\rm syn,pk}$ and
$L_{\nu_{\rm syn,pk}}$ are the synchrotron peak frequency and specific luminosity.\footnote{Here and hereafter we use the convention $Q_x=\left(Q/10^x\right)$ in cgs units, except for the BH mass, $M = 10^x M_x M_\odot$.}
We fix the spectral index of seed photons as $p=2$, since the dependence of the gap dynamics on $p$ is minor (see \ref{subsec:weak}) and the spectral slope of thermal component around the peak is not too much deviated from $E_\gamma F_{E_\gamma}\propto E_\gamma^{-1}$.
We do not consider reduction of the mass accretion rate due to the disk wind for simplicity.
The magnetic field strength at the BH horizon $\Bh$ is estimated by Eq.~(\ref{bh}).

In Fig.~\ref{Mnp} we demonstrate the resultant color maps of $L_{\rm cur,pk}/\Lbz$ calculated for $3.2M_\odot\leq M\leq 10^3M_\odot$ and $1.0\,{\rm cm}^{-3}\leq  n_{\rm ism}\leq 10^4\,{\rm cm}^{-3}$ in the cases of $j_0 = -1/2\pi$ and $-1/2$. 
As can be seen, $L_{\rm cur,pk}/\Lbz$ is higher in the parameter regions of smaller $M$ and smaller $n_{\rm ISM}$ for both of the cases of $j_0$. (Note that $\Lbz$ has the opposite trend.)
Combining the results of the two cases of $j_0$, we find $L_{\rm cur,pk}/\Lbz \gtrsim 10^{-4}$ for a limited range of $\nism$: 
For $M \sim 10 M_\odot$, the range is $50\,\mathrm{cm^{-3}}\lesssim\nism\lesssim3.0\times10^2\,\mathrm{cm^{-3}}$, which corresponds to dense molecular clouds, while for $M \sim 50 M_\odot$, the range is $5\,\mathrm{cm^{-3}}\lesssim\nism\lesssim30\,\mathrm{cm^{-3}}$, which corresponds to cold neutral medium.
The parameters for the white regions in Fig.~\ref{Mnp} yield $\tau_0\lesssim27$ for the $j_0=-1/2\pi$ case and $\tau_0\lesssim14$ for the $j_0=-1/2$ case, for which we could not find the solution of the gap in $r>r_{\rm in}$.
The gap would achieve its maximum efficiency of particle acceleration when $\tau_0$ is just above this critical value, and correspondingly, the curvature luminosity hits the highest.
These critical values of $\tau_0$ are consistent with our simulation results:
The simulation box slowly approaches the vacuum state for models of $\tau_0 = 10$ with all the three $j_0$ values (though not shown in Table~\ref{tab:models}), similarly to what we saw in the SMBH cases.\footnote{Our simulations do not include contribution of pairs created by curvature photons, which could slightly alter the critical $\tau_0$ value (see also K20).}

\begin{figure*}[ht!]
\hspace*{-2.3cm}
\centering
\begin{minipage}[b]{0.45\linewidth}
 \includegraphics[keepaspectratio, trim = 5 20 100 0, scale=0.48]{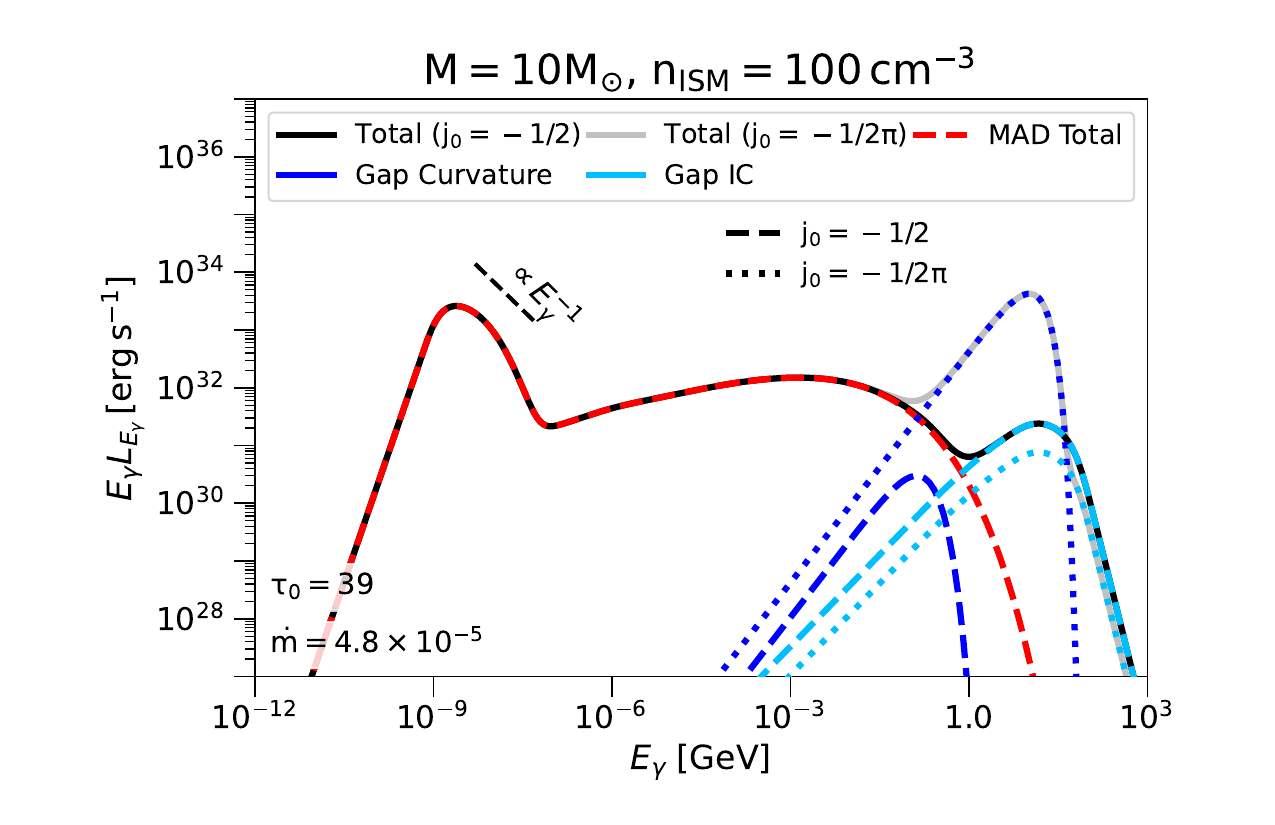}
 \end{minipage}
 \hspace*{0.01\columnwidth}
 \begin{minipage}[b]{0.45\linewidth}
 \includegraphics[keepaspectratio, trim = 5 20 100 0, scale=0.48]{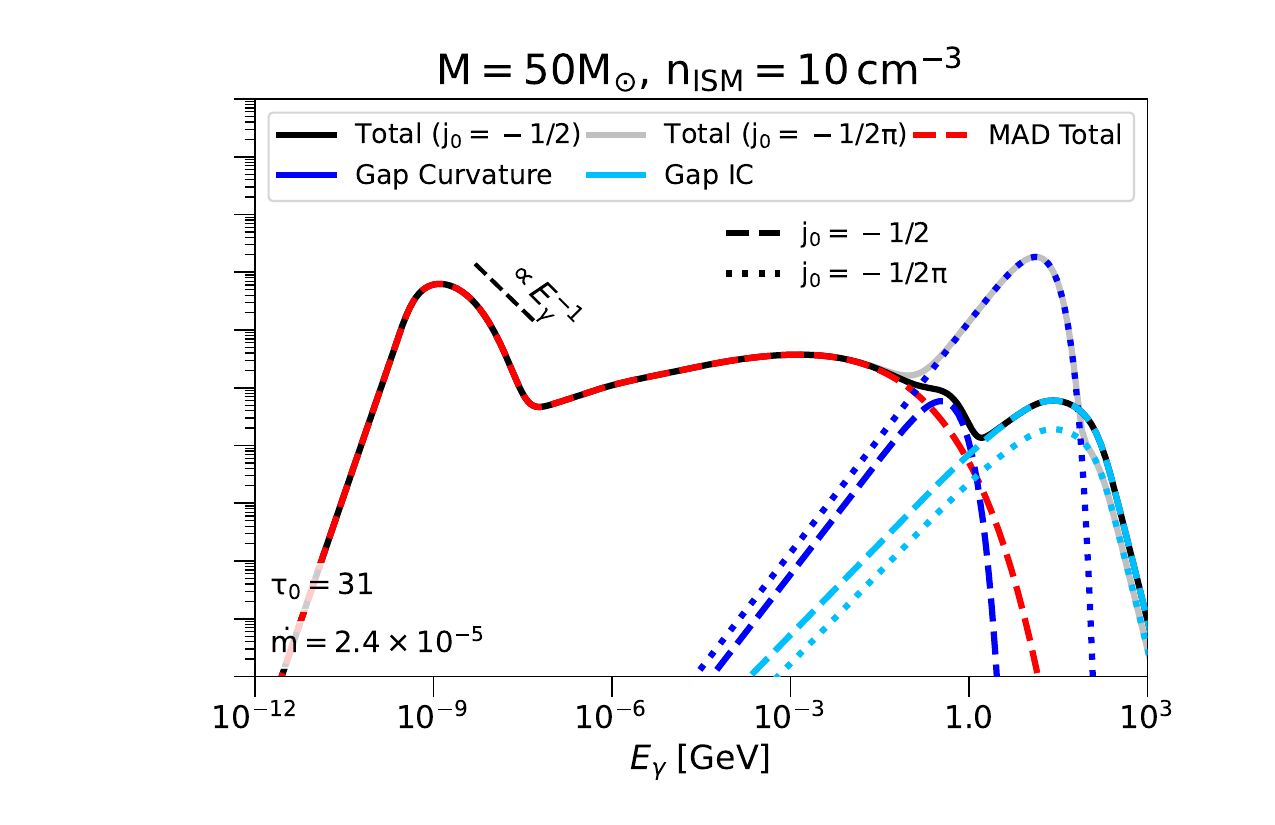}
 \end{minipage}
\caption{
Broadband emission spectra from IBH-MADs with magnetospheric gaps in cases of $M = 10M_\odot,\,n_{\rm ISM} = 100~{\rm cm}^{-3}$ (\textit{left panel}) and  $M = 50M_\odot,\,n_{\rm ISM} = 10~{\rm cm}^{-3}$ (\textit{right panel}). 
The value of $\tau_0$ and $\dot{m}=\dot{M}/\dot{M}_{\rm Edd}$ are shown in each panel.
The black and gray solid lines represent the total emission for $j_0 = -1/2$ and $j_0=-1/2\pi$, respectively. 
The red dashed line represents radiation from the MAD, which consists of synchrotron radiation from the MAD thermal electrons (which forms the humps in the optical band) and that from the MAD non-thermal electrons.
Spectra of the curvature and IC radiation from the gap are shown by the blue and light-blue lines, respectively.
For each of them, the dashed lines correspond to those for $j_0=-1/2$ case and the dotted ones for $j_0=-1/2\pi$ case.}
\label{spectrum}
\end{figure*}
We present examples of broadband emission spectra from IBH-MADs with maghetospheric gaps based on our semi-analytic calculations in Fig.~\ref{spectrum}.
The parameters for these examples are marked by the black stars in Fig.~\ref{Mnp}.
The spectral components from the magnetospheric gaps and the MADs, taking into account the internal attenuation by $\gamma\gamma$ pair annihilation of gamma rays and MAD photons, are calculated by the methods in Appendix~\ref{specdetail} and \citet{Kimura21}, respectively.
One must be noted that we consider time-averaged luminosities when calculating gap emissions:
We observe $L_{\rm cur}$ and $L_{\rm ic}$ oscillating in the simulations (Fig.~\ref{lc}).
The oscillation period $T_{\rm gap}\lesssim10\rg/c\simeq5.0\times10^{-4}M_1{\rm s}$, however, cannot be resolved by ongoing or forthcoming gamma-ray detectors for stellar-mass BHs.
Hence, we calculate the observed spectral luminosities by reducing the intrinsic peak values by the typical duty cycle seen in the simulations, $60\%$.
As seen in Fig.~\ref{spectrum}, the gap emission peaks in the GeV-TeV energy band for the both of $j_0=-1/2\pi$ and $-1/2$.
\begin{figure}
\centering
 \includegraphics[keepaspectratio, trim = 120 0 300 0, scale=0.55]{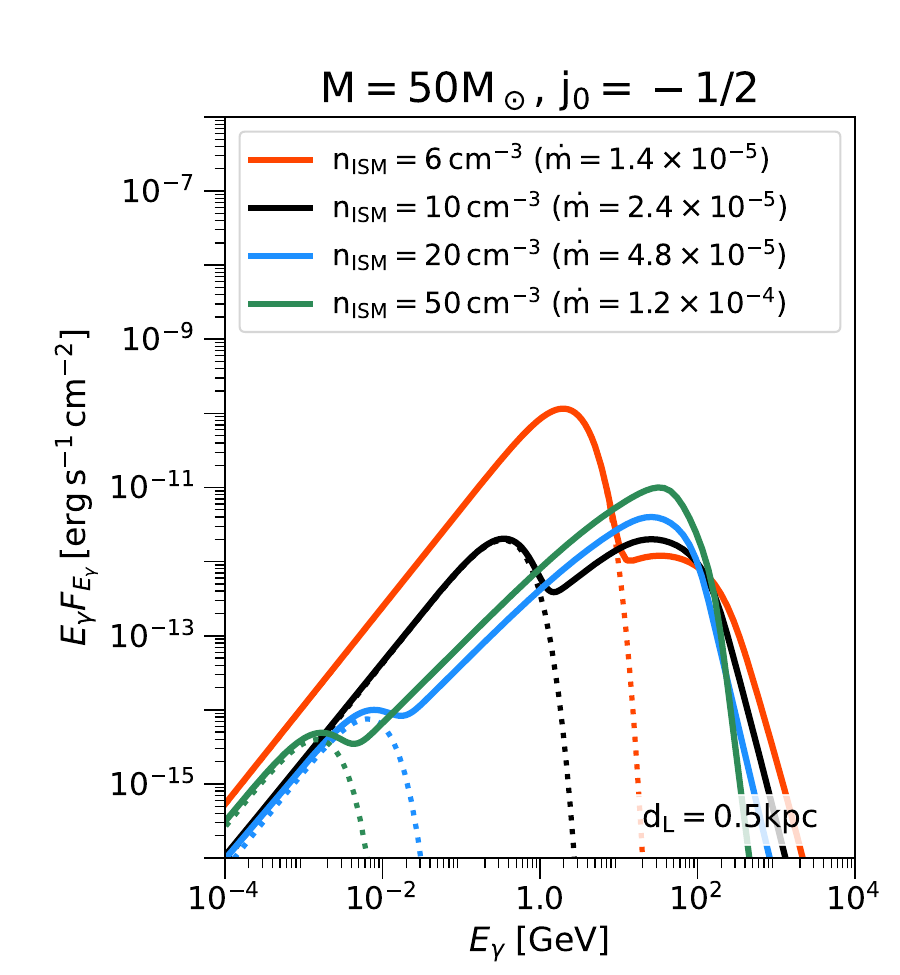}
 \hspace*{4.0cm}
\caption{Observed gap gamma-ray fluxes for $M =50M_\odot$, $j_0=-1/2$, $d_{\rm L}=0.5\,\rm kpc$. The solid lines indicates the total spectra for $\nism=6\,\rm cm^{-3}$  (\textit{red}), $10\,\rm  cm^{-3}$ (\textit{black}), $20\,\rm  cm^{-3}$ (\textit{light-blue}), and $50\,\rm  cm^{-3}$ (\textit{green}), respectively. The curvature emission components are shown with the dotted lines.}
\label{spec_ncomp}
\end{figure}
The spectral cutoff at the sub-TeV energy band is the result of the pair annihilation with synchrotron photons from the MAD thermal electrons (see Appendix~\ref{specdetail}).
For a given set of $M$ and $\nism$, the spectral luminosity in the GeV-TeV energy band is higher for $j_0 = -1/2\pi$, owing to the efficient curvature radiation.
For $j_0=-1/2$, the curvature peak luminosity and the characteristic frequency are $10^2-10^3$ times lower than those for $j_0=-1/2\pi$, but IC emission is dominant in the GeV-TeV energy band due to the scaling $L_{\rm ic,pk}\propto|j_0|$ (see \ref{subsec:j0p5}).
As a result, the gamma-ray signals from IBH magnetospheric gaps would be as bright as $L\gtrsim10^{-4}\Lbz$ over wide ranges of $M$ and $\nism$.

For an illustrative purpose, we show the gamma-ray fluxes
from gaps of a nearby ($d_{\rm L}=0.5\,\rm kpc$) IBH of $M=50 M_\odot$ with various values of $\nism$ in Fig.~\ref{spec_ncomp}.
The peak fluxes exceeds $E_{\gamma}F_{ E_\gamma}\sim10^{-12}\,\rm erg\, s^{-1}cm^{-2}$, which is high enough for detection with current gamma-ray detectors, including Fermi-LAT \citep{F-LAT}.

\section{Summary and Discussion}\label{sec:dandc}
We have conducted a comprehensive investigation of the dynamics of spark gaps in charge-starved magnetospheres of stellar-mass Kerr BHs, using the 1D GRPIC code described in \citet{Levinson18}, K20, and K22.
Differences of dynamics for changing spectral properties of seed photons and the magnetospheric current have been examined.
We have also presented the semi-analytic model that reproduces maximum electron Lorentz factors and peak gamma-ray luminosities of the simulation results for various parameter values.
Using this model, we have estimated the characteristics of gamma-ray signals from IBHs over broad ranges of BH masses and infalling ISM densities.
The gamma-ray signals from gaps of IBHs embedded in the dense gas clouds located at the distances $\lesssim\rm\,kpc$ are suggested to be bright enough for detections in the GeV energy band.

\subsection{Additional Pair Production Processes in the Magnetosphere}\label{ss:otherpair}
In this work we have assumed the absence of sufficient plasma injection into the magnetosphere and considered only the pair production process by annihilation of IC scattered photons and soft photons from the MAD. Here we discuss other pair production processes in the magnetosphere that might screen the gap.
One possibility we need to consider is the steady plasma injection via annihilation of MeV photons from the MAD \citep{Levinson11,Kimura20}.
The expected pair number density around the null surface is 
$n_\pm\sim(r_{\rm null}\sigma_{\gamma\gamma}/16\pi^2R^4c^2)\left(L_{\rm MeV}/E_{\rm MeV}\right)^2\simeq10^5\;\mathcal{R}_1^{-4}M_1n_{\rm ISM,2}^2\,\rm cm^{-3}$, where $L_{\rm MeV}\simeq10^{32}M_1^2 n_{\rm ISM,2}\,\rm erg\,s^{-1}$ and $E_{\rm MeV}$ are the spectral luminosity (see Fig.~\ref{spectrum}) and the characteristic energy in the MeV energy band, respectively.
This is much lower than the GJ density $n_{\rm GJ}(\rnull)\sim\Omega_{\rm F}B(\rnull)/2\pi ec\simeq
1.4\times10^8\, (\phi_{\rm BH}/50)M_1^{-1}n_{\rm ISM,2}^{1/2} \,{\rm cm}^{-3}$, where $B(\rnull)\sim\Bh(\rnull/\rh)^{-2}\sim\Bh/2$ is the magnetic field strength around the null charge surface.
Furthermore, $n_\pm/n_{\rm GJ}$ and $\tau_0$ have the same dependence, $\propto M^2\nism^{3/2}$, hence the MeV photon annihilation will not inject sufficient amount of pairs to screen the gap for ranges of $M$ and $\nism$ for which gamma-ray radiation is bright.

Alternatively, as briefly mentioned in Section~\ref{sec:intro}, sporadic flares by magnetic reconnection around the equatorial plane at $r_{\rm rec}\sim2\rg$ can produce MeV photons in the magnetosphere \citep{Chashkina21, Ripperda22, Kimura22, Chen22, Hakobyan23}.
During each flare, the magnetic energy is dissipated in a localized reconnection region of a size $l_{\rm rec}\sim\rg$ with the reconnection velocity $\beta_{\rm rec} \sim 0.1$ \citep[e.g.,][]{Guo20}, and the accelerated electrons are efficiently cooled by emitting MeV photons.
Then the luminosity is $L_{\rm MeV}\sim 2(B(r_{\rm rec})^2/8\pi)l_{\rm rec}^2\beta_{\rm rec}c\sim(\phi_{\rm BH}^2/64\pi^3)\beta_{\rm rec}\dot{M}c^2\simeq
8\times 10^{33}(\phi_{\rm BH}/50)^2\beta_{\rm rec,-1}M_1^2n_{\rm ISM,2}\,\rm erg\,s^{-1}$.
Thus, one obtains $n_\pm\simeq6\times10^{12}(d/\rg)^{-4}(\phi_{\rm BH}/50)^4 \beta_{\rm rec,-1}^2 M_1^5$
$n_{\rm ISM,2}^2\rm\, cm^{-3}$ in a similar manner to the above discussion for MAD MeV photons, replacing $R$ with the distance between the emitting region and the gap $d\sim\rg$.
This is $\sim10^5$ times higher than $n_{\rm GJ}(\rnull)$, so that the gap will be completely screened during the flare. 
\citet[][]{Ripperda22} suggests that each flare is followed by a quiescent phase of timescale $\sim10^3\rg/c$.
The injected pairs will escape from the null surface in a timescale $\sim\rg/c$ after the flare, and hence, the gap is expected to develop in the quiescent phase.

A possibility of single photon pair production, or magnetic pair production, in the magnetosphere also needs to be considered.
Its opacity can be written as functions of the interacting photon energy $\epsilon_\gamma$, the local magnetic field $B$, and the angle between the photon momentum and the magnetic field $\psi$ \citep{Erber66},
\begin{equation}
\label{gBop}
\alpha_B(\epsilon_\gamma,\,B,\,\psi)\approx0.23\dfrac{\alpha_f}{\lambdabar_c}b\sin\psi\exp\left(-\dfrac{4}{3\chi}\right),
\end{equation}
where $\alpha_f=e^2/\hbar c\simeq1/137$ is the fine structure constant, $\lambdabar_{\rm c}=\hbar/\me c\simeq3.9\times10^{-11}\rm cm$ is the reduced Compton wavelength, $b=B/B_q$ is the local magnetic field strength normalized by $B_q \equiv e/\alpha_f\lambdabar_{\rm c}^2\simeq4.4\times10^{13}\rm G$, and $\chi=(1/2)\epsilon_\gamma b\sin\psi$.
$B(\rnull)\simeq5\times10^6B_{\rm H,7}\rm G$ for $M\sim10M_\odot$ and $\nism\sim10^2\rm cm^{-3}$, for which the gamma-ray is bright and the magnetic field is strong (see Fig.~\ref{Mnp} and the left panel of Fig.~\ref{spectrum}).
The particle Lorentz factor in this case is $\gammae\sim10^7$, and thus the photons produced in the gap via IC scattering will be $\epsilon_{\gamma}\sim3.0\times10^6$ (see Eq.~(\ref{eic})).\footnote{We can exclude the possibility of the magnetic pair production for curvature photons because of their much lower typical energy $\epsilon_{\rm cur}\sim10^4$ (see Eq.~\ref{ecur}).}
The corresponding optical depth of the magnetic pair production is $\tau_B\sim\alpha_B\rg\simeq4\times10^3$ for $\sin\psi=1.0$, and $2\times10^{-2}$ for $\sin\psi=0.4$. In the latter estimate we have approximated $\psi$ by the field pitch angle $|B_\phi|/B_r \sim 0.4$ measured in the ZAMO frame \citep[see Appendix A in][]{Kimura22}.
Such a sensitive dependence of $\tau_B$ on $\psi$ implies that the detailed magnetic field structure and the general relativistic lensing effect significantly alter the pair loading efficiency. 
To analyze them is beyond the scope of this paper.
Nevertheless, we confirmed that magnetic pair production is unimportant for models shown in the right panel of Fig.~\ref{spectrum} and all the models in Fig.~\ref{spec_ncomp}, regardless of $\psi$:
For smaller $\nism$, the magnetic field is weaker.
For $\tau_0\propto\nism^{3/2}\gg30$, $\epsilon_\gamma\lesssim10^6$ and the gap width is much smaller than $\rg$.

\subsection{Possible Strategy for Candidate Search}
Our expected main targets are Fermi-LAT unidentified objects (unIDs).
The catalog from 12-year operations of Fermi-LAT \citep[4FGL-DR3;][]{4FGL-DR3} contains more than $2000$ objects lacking associations with known pulsars/supernovae/AGNs.
Nearly a half of them are lying at a low galactic latitude ($|b|<10$), aligning with our expectation of molecular cloud associations.
Our IBH candidates can be distinguished by a hard spectral index ($dN/dE_\gamma\propto E_\gamma^{-\Gamma}$ ($\Gamma\sim2/3$))
with a break around the GeV-TeV and a point-source like morphology.
Since objects with a hard photon index are rare in the entire population of unIDs, we may be able to narrow down the candidates solely by the Fermi-LAT data.
Furthermore, we can identify IBHs by taking correlations with survey catalogs of other wavelength.
Our model also predicts the bright radiation from the MADs at optical-to-MeV band (see Fig.~\ref{spectrum}; \citealp{Kimura21}).
The optical signals from the MADs resembles that from white dwarfs, which can be found by Gaia \citep{Gaia}, and the MAD X-ray signals can be detected by X-ray observations including eROSITA \citep{eROSITA} and ROSAT \citep{ROSAT}.
Thus, we could specify our IBH candidates by examining associations among Fermi-LAT unIDs, X-ray sources, and Gaia white-dwarf-like objects.

Time variations of the gamma-ray luminosity due to a fluctuation in the mass accretion rate can also be useful to distinguish IBHs from other objects.
$L_{\rm cur,pk}\propto\tau_0^{-8/3}\Lbz\propto\dot{M}^{-3}(\propto\nism^{-3})$ and $L_{\rm ic,pk}\propto\Lbz\propto\dot{M}$, hence occasional luminosity variations by a factor of $\lesssim 10^2$ (as shown in Fig.~\ref{spec_ncomp}) are expected.
The timescale of such variations would be no shorter than the interaction timescale of IBHs and ISM clouds, $T_{\rm var}\sim GM/v^3\simeq2.1\times10^7M_1(v/40\,{\rm km\,s^{-1}})^{-3}\,\rm s$, where $v\simeq40\,\rm km\,s^{-1}$ is the IBH proper motion velocity.
Multi-epoch observations with CTA \citep{CTA} can be utilized to capture luminosity variations in such cases.
Fermi-LAT can also detect the gamma-ray variability \citep{Abdollahl17, 4FGL} if the source is bright.

\subsection{Model Comparison and Uncertainty}
\citet{Hirotani16_2}, \citet{Hirotani17}, \citet{Song17}, and \citet{Hirotani18} also examined the gap GeV-TeV radiation for various masses of BHs with their analytical 2D gap models.
They assume that the gap is steadily maintained.
However, various numerical simulations have shown that the gap is time-variable, and our model reproduces the characteristics of gap at the peak time.
They obtain non-zero charge density in the gap, while we assume the vacuum solution of the electric field, consistent with the simulation results. 
Such differences of model settings and assumptions result in various differences of calculations results:
The IC component in their calculations peaks at a higher energy range with higher luminosity than our model.
Furthermore, the luminosities do not significantly depend on the magnetospheric current in their model \citep[see Figure~6 of][]{Hirotani18}.

It is uncertain whether local stellar-mass BHs can acquire a high spin.
One possibility is the spin-up via a binary BH merger \citep{Zlochower15}, but it would be still challenging to achieve an extremely high spin rate as $a=0.9$.
We thus need to take into account the fraction of rapidly-rotating stellar-mass BHs when discussing the expected number of IBH detections.
In addition, it would be important to understand the BH spin dependence of the gap dynamics and radiation features.

For calculating the spectrum of IC emission escaping from the soft photon field in the magnetosphere ($r\leq R=10\rg$), we have assumed that $L_{\rm ic,pk} \sim$ const. outside the gap by neglecting synchrotron loss (Section~\ref{sec:impli} and Appendix~\ref{sec:ic}). 
However, secondary pairs have momenta in the direction of incident IC photons, which is not necessarily aligned with the local magnetic field vector,
so that they could lose their momenta perpendicular to the local field via synchrotron emission.
Indeed, for secondary pairs with the finite pitch angle $\alpha$ and the Lorentz factor $\gamma_{\rm sec} \sim\epsilon_{\rm ic}/2\sim10^6$, we have $t_{\rm cool}^{\rm syn}\ll\tcic$.\footnote{We might need to take account of a quantum reduction of the synchrotron power for $\gamma_{\rm sec}\sim10^6$ particles \citep[c.f.][]{Erber66, Harding06}, which is effective for $\gammae (B/B_q)\sin\psi>0.1$.}
The synchrotron photon's typical energy is $\epsilon_{\rm syn}=(3heB/4\pi\me^2c^3)\gamma_{\rm sec}^2\sin\alpha\simeq1.7\times10^5\sin\alpha (B/5.1\times10^6\rm G)\gamma_{{\rm sec},6}^2$, thus they will have large mean free paths for pair annihilation $l/R\sim\tau_0^{-1}(\epsilon_{\rm syn}^{-1}/\emin)^2\sim 1.2 (\sin\alpha)^{-2}\epsilon_{\rm min,-6}^{-2}(B/5.1\times10^6\rm G)^{-2}\gamma_{{\rm sec},6}^{-4}(\tau_0/30)^{-1}$.
Since $\sin\alpha<1$, most of the synchrotron photons may escape from the region of $r<R$.
Since $\epsilon_{\rm syn}$ is also in the GeV-TeV energy band, the peak luminosity of the IC plus synchrotron radiation may not substantially change from our rough estimate of IC luminosity in Section~\ref{sec:impli}.
Detailed consideration of multi-dimensional magnetic field structure and $\alpha$ would be required to calculate the IC and synchrotron spectra more accurately.

Our 1D calculations do not take account of a general relativistic lensing effect on the photon transfer.
This affects the light curve, which will be considered in future work.
Differences between 1D and 2D GRPIC simulation results on the gap dynamics should also be further investigated.
We have shown that $J_0$ (or the toroidal magnetic field) substantially affects the gap dynamics. The toroidal field is determined by external pressure on the magnetosphere. Including effects of variable external pressure in 2D simulations would be interesting.
\\
\\
We thank the anonymous referee for useful comments.
Numerical computations in this research are performed on Cray XC50 at Center for Computational Astrophysics of National Astronomical Observatory of Japan, and Yukawa-21 at Yukawa Institute for Theoretical Physics in Kyoto University.
We utilized Science Lounge of FRIS CoRE for discussions many times.
This work was supported by Graduate Program on Physics for the Universe (GP-PU), Tohoku University (KK), JST SPRING, Grant Number JPMJSP2114 (KK), a grant from the Simons Foundation (00001470, AL), and Japan Society for the Promotion of Science Grants-in-Aid for Scientific Research (KAKENHI) Grant Numbers JP21H01078 (SK), JP22H01267 (SK), JP22K03681 (SK), 22K14028 (SSK), 21H04487 (SSK), 23H04899 (SSK).
SSK acknowledge the support by Tohoku Initiative for Fostering Global Researchers for Interdisciplinary Sciences (TI-FRIS) of MEXT’s Strategic Professional Development Program for Young Researchers.

\appendix
\section{Spacetime}\label{space}
The Kerr metric is given in Boyer-Lindquist coordinates as
\begin{equation}
ds^2=-\alpha^2dt^2+g_{\phi\phi}\left(d\phi-\omega dt\right)^2+g_{\rm rr}dr^2+g_{\theta\theta}d\theta^2,
\end{equation}
where
\begin{equation}
\alpha^2=\dfrac{\Sigma\Delta}{A}, ~~~\omega=\dfrac{2ar_gr}{A}, ~~~g_{rr}=\dfrac{\Sigma}{\Delta}, ~~~g_{\theta\theta}=\Sigma, ~~~g_{\phi\phi}=\dfrac{A}{\Sigma}\sin^2\theta,
\end{equation}
with
\begin{equation}
\Delta=r^2+a^2-2r_gr, ~~~\Sigma=r^2+a^2\cos^2\theta, ~~~A=\left(r^2+a^2\right)^2-a^2\Delta\sin^2\theta.
\end{equation}
In our simulations, the radial coordinate is replaced by the tortoise coordinate $\xi = \rg\ln[(r-r_+)/(r-r_-)]/(r_+ - r_-)$, where $r_{\pm} = r_g(1 \pm \sqrt{1-a_*^2})$.
We note that $\xi\rightarrow-\infty$ as $r\rightarrow\rh$, and $\xi\rightarrow0$ as $r\rightarrow\infty$. 
The simulating region $-3 \leq \xi \leq -0.3$ corresponds to $1.5\rg\lesssim r\lesssim4.3\rg$.

\section{Spectra of curvature and IC emissions}\label{specdetail}
\subsection{Curvature emission}
We calculate the observed spectrum of curvature emission from the magnetospheric gaps at the peak time, for which we may assume that the emitting electrons have mono-energetic energy distribution $\gammae \approx \gamma_{\rm e,max}$. 
The radiated power of one electron with Lorentz factor $\gammae$ per unit energy is
\begin{equation}
P_{\epsilon}^{\rm cur}(\gammae)=\dfrac{\sqrt{3}e^2\gammae}{h\rg}F\left(\dfrac{\epsilon}{\epsilon_{\rm cur}}\right),
\end{equation}
where
\begin{equation}
F(x)=x\int^{\infty}_{x}d\xi K_{5/3}(\xi),
\end{equation}
$K_{5/3}(\xi)$ is the modified Bessel function of $5/3$ order, and
\begin{equation}
\label{ecur}
\epsilon_{\rm cur}=\dfrac{3}{4\pi}\dfrac{hc}{\rg}\gammae^3
\end{equation}
is the characteristic energy \citep{radipro}.
Let $N_e^{\rm cur}$ denote the number of electrons with $\gammae \approx \gamma_{\rm  e,max}$, then we can derive the specific luminosity as
\begin{equation}
\begin{aligned}
L_{\epsilon}^{\rm cur}=\dfrac{\sqrt{3}e^2\gamma_{\rm e,max}}{h\rg}N_e^{\rm cur}F\left(\dfrac{\epsilon}{\epsilon_{\rm cur}(\gamma_{\rm e,max})}\right).
 \end{aligned}
 \end{equation}
$N_e^{\rm cur}$ is determined so that $L_{\rm cur,pk} = \int^{\infty}_0L_{\epsilon}^{\rm cur}d\epsilon$.
The attenuation of gamma-rays via pair annihilation is evaluated by multiplying $L_{\epsilon}^{\rm cur}$ by the factor $\exp{[-\tau_{\gamma\gamma}(\epsilon,\,\epsilon_{\rm s})}]$.
Here $\tau_{\gamma\gamma}$ is the pair annihilation optical depth (see, e.g., \citealt{Kimura19} and references therein) at $r=R$, within which the seed photons are dense ($R=10\rg$ is the size of the MAD (see Section~\ref{sec:impli})).
We show only the attenuated spectrum in Fig.~\ref{spectrum}.
The overall spectrum is gravitationally redshifted by $\alpha(r_{\rm emit})$, where we set the emission radius of curvature photons to be $r_{\rm emit}=r_{\rm null}$. 

\subsection{IC emission}
\label{sec:ic}
For $\tau_0 \gtrsim 10$, for which the magnetosphere does not approach the vacuum, the primary IC photons produced by electrons with $\gammae \approx \gamma_{\rm e,max}$ annihilate with the seed photons to create $e^+e^-$ pairs in the gap.
These pairs produce gamma-rays via IC scatterings even outside the gap. Here we do not solve the detailed radiation transfer, but simply estimate the spectrum of the IC emission escaping from $r=R$ of the magnetosphere.

The IC gamma-rays induce the pair cascade while propagating in the magnetosphere. (Note that we do not consider the cascade of IC gamma-rays injected into the MAD.) We find $\tcic \ll \tdyn \ll \tccur$ outside the gap, where $\tcic$, $\tccur$, and $\tdyn$ are the IC cooling timescale, the curvature cooling timescale, and the dynamical timescale, respectively.
The particle kinetic luminosity thus keeps sub-dominant outside the gap as shown in Fig.~\ref{lc}, and the energy loss via curvature emission is negligible, so that one may assume $L_{\rm ic,pk} \sim {\rm const.}$ outside the gap.
The simulation results indicate $L_{\rm ic,pk}\sim10^{-3}|j_0|\Lbz$.
The characteristic energy of escaping IC photons $\epsilon_{\rm ic,esc}$ can be estimated as the energy for which the pair annihilation optical depth $\tau_{\gamma\gamma}(\epsilon_{\rm ic,esc},\,\epsilon_{\rm s})=1$ (see \citealp{Timokhin15, Kisaka17}; K20).
The Lorentz factor of electrons that produce the IC photons of energy $\epsilon_{\rm ic, esc}$ is evaluated as $\gamma_{\rm ic,esc}=\min\{\gamma_{\rm e,max},\,\sqrt{\epsilon_{\rm ic,esc}/0.3\emin}\}$.

The spectral luminosity is then given by
\begin{equation}
\begin{aligned}
&L_\epsilon^{\rm ic}=4\sT N_e^{\rm ic}\int^{1}_{0}dx\,g(x)F_{\epsilon_{\rm s}}(x),\hspace{20pt}x = \dfrac{\epsilon}{4\gamma_{\rm ic,esc}^2\epsilon_{\rm s}},
\end{aligned}
\end{equation}
where $g(x)=1+x+2x\ln{x}-2x^2$ is a kernel in the relativistic limit \citep{Blumenthal70,radipro} and $F_{\epsilon_{\rm s}}(x)$ denotes the seed photon flux.
$N_e^{\rm ic}$ is determined to satisfy $L_{\rm ic,pk} = \int^{\infty}_0L_{\epsilon}^{\rm ic}d\epsilon$.
To evaluate the attenuation at $\epsilon>\epsilon_{\rm ic,esc}$, we multiply $L_{\epsilon}^{\rm ic}$ by $[1-\exp{(-\tau_{\gamma\gamma}(\epsilon,\,\epsilon_{\rm s})})]/\tau_{\gamma\gamma}$.
We show only the attenuated spectrum in Fig.~\ref{spectrum}.
The spectrum is gravitationally redshifted by $\alpha(r_{\rm emit})$, where we set $r_{\rm emit}=R$. 


\bibliography{sample631}{}
\bibliographystyle{aasjournal}



\end{document}